\documentclass[apj]{emulateapj}
\usepackage{multirow}
\usepackage{subfigure}

\def\xmm{{\it XMM-Newton}}

\shortauthors{Lin et al.}
\begin{document}
\title{Classification of X-ray Sources in the {\it XMM-Newton} Serendipitous Source Catalog}

\author{Dacheng Lin\altaffilmark{1,2}, Natalie A. Webb\altaffilmark{1,2}, Didier Barret\altaffilmark{1,2}}
\altaffiltext{1}{CNRS, IRAP, 9 avenue du Colonel Roche, BP 44346, F-31028 Toulouse Cedex 4, France, email: Dacheng.Lin@irap.omp.eu}
\altaffiltext{2}{Universit\'{e} de Toulouse, UPS-OMP, IRAP, Toulouse, France}

\begin{abstract}
We carry out classification of 4330 X-ray sources in the 2XMMi-DR3
catalog. They are selected under the requirement of being a point
source with multiple \textit{XMM-Newton} observations and at least one
detection with the signal-to-noise ratio larger than 20. For about one
third of them we are able to obtain reliable source types from the
literature. They mostly correspond to various types of stars (611),
active galactic nuclei (AGN, 753) and compact object systems (138)
containing white dwarfs, neutron stars, and stellar-mass black
holes. We find that about 99\% of stars can be separated from other
source types based on their low X-ray-to-IR flux ratios and frequent
X-ray flares. AGN have remarkably similar X-ray spectra, with the
power-law photon index centered around 1.91$\pm$0.31, and their
0.2--4.5 keV flux long-term variation factors have a median of 1.48
and 98.5\% less than 10. In contrast, 70\% of compact object systems
can be very soft or hard, highly variable in X-rays, and/or have very
large X-ray-to-IR flux ratios, separating them from AGN. Using these
results, we derive a source type classification scheme to classify the
other sources and find 644 candidate stars, 1376 candidate AGN and 202
candidate compact object systems, whose false identification
probabilities are estimated to be about 1\%, 3\% and 18\%,
respectively. There are still 320 associated with nearby galaxies and
151 in the Galactic plane, which we expect to be mostly compact object
systems or background AGN. We also have 100 candidate ultra-luminous
X-ray sources. They are found to be much less variable than other
accreting compact objects.
\end{abstract}


\section{INTRODUCTION}
\label{sec:intro}
The \xmm\ observatory has been carrying out pointed observations of
the X-ray sky for more than a decade. The \xmm\ Serendipitous Source
Catalog \citep{wascfy2009} provides various properties of
serendipitously detected X-ray sources, in addition to targets, from
these pointed observations. It has been updated regularly, and the
latest version as of the writing of this paper is the 2XMMi-DR3
catalog\footnote{http://xmmssc-www.star.le.ac.uk/Catalogue/xcat\_public\_2XMMi-DR3.html},
containing 353191 X-ray source detections for 262902 unique X-ray
sources. It is the largest X-ray source catalog ever produced.

Such catalogs are rich resources for exploring a variety of X-ray
source populations and identifying rare source types
\citep[e.g.,][]{licagr2011,pimoca2011,pimoja2009,faweba2009}. Compared
with other large X-ray catalogs such as the {\it ROSAT} catalogs from
its all-sky survey (RASS) and pointed observations and the Chandra
Source Catalog \citep{whgian1994,voasbo1999,voasbo2000, evprgl2010},
the 2XMMi-DR3 catalog has several advantages because of the large
effective area, high spatial resolution, and/or broad energy band
coverage of the \xmm\ observatory.

In this paper, we select a sample of sources of the best quality from
the 2XMMi-DR3 catalog and study their properties in detail. We first
establish a subsample of these sources whose types are known and study
their X-ray spectral shapes, X-ray variability, and X-ray-to-optical
and X-ray-to-IR flux ratios. Based on these properties, we define some
quantitative criteria to classify other sources. Sources with
interesting properties, discovered as a part of this work are the
subject of a companion paper. We devote significant effort to visually
screen the results so as to reduce various kinds of systematic errors.

In Section~\ref{sec:reduction}, we describe the source selection
procedure, the calculation of the flux and the hardness ratio, the
search for stellar X-ray flares, the measurement of the long-term
variability, the source classification from the literature, the search
for optical and IR counterparts, the determination of the
(extra)galactic nature, and simple spectral fits. The properites of
the sources with known types are shown in Section~\ref{sec:res}. The
scheme that we propose to classify sources is given in
Section~\ref{sec:clsscrit}. The results of applying this scheme to the
sources with unknown types are presented in
Section~\ref{sec:srcuid}. The discussion and conclusions of our study
are given in Section~\ref{sec:conclusion}.

\section{DATA ANALYSIS}
\label{sec:reduction}

\subsection{The 2XMMi-DR3 Catalog and Source Selection}
The 2XMMi-DR3 catalog is based on 4953 pointed observations. These
observations have a variety of modes of data acquisition, exposures
($>$1000 s), and optical blocking filters (thick, medium, and
thin). The catalog contains detections in five basic individual energy
bands (0.2--0.5, 0.5--1.0, 1.0--2.0, 2.0--4.5, and 4.5--12.0 keV,
numbered 1--5, respectively), for each European Photon Imaging Camera
(EPIC), i.e., pn, MOS1/M1, and MOS2/M2
\citep{jalual2001,stbrde2001,tuabar2001}. Results are also given for
various combinations of energy bands and cameras. Band 8 refers to the
total band 0.2--12.0 keV, and EPIC/EP refers to all cameras. We also
follow these notations throughout this paper.

We focused only on point sources with multiple \xmm\ observations and
at least one detection with the signal-to-noise ratio S/N $\ge$20 in
this paper. We considered only point sources because extended sources
(typically Galactic supernova remnants (SNRs) and galaxy clusters)
tend to be subject to complicated problems of source detection and
flux measurement \citep{wascfy2009}. We define a source to be
point-like when its extent (\verb|EP_EXTENT| in the catalog) is zero
(at least in one detection). We note that in the pipeline, if the
extent radius in a detection is $<$6$\arcsec$, it is set to zero,
i.e., a point source is assumed. We estimated the S/N using
\verb|EP_8_CTS|/\verb|EP_8_CTS_ERR|, where \verb|EP_8_CTS|, given in
the catalog, is the total counts summed over all cameras, and
\verb|EP_8_CTS_ERR| is its error. Our requirement of having at least
one detection with S/N $\ge$20 is to ensure at least one good spectrum
for detailed analysis when needed. The requirement of having multiple
observations allows us to calculate the long-term variability. It is
different from the requirement of having multiple detections. The
catalog includes only detections with a total likelihood (from all
bands and all cameras) $>$ 6 (corresponding to about $3\sigma$). We
used the service \verb|http://www.ledas.ac.uk/flix/flix.html| to find
the observations of a source and estimate the fluxes for the
observations that have no entries in the catalog.

There are 5802 sources satisfying the above conditions. However, as
the catalog was produced using an automated procedure, some of these
sources suffer from various kinds of problems, and we excluded them
based on visual inspection. About 16\% are spurious, mostly due to
single reflections caused by bright sources outside the field of view,
out-of-time events, and extremely bright sources
\citep{wascfy2009}. We also excluded about 8\% of the sources that are
detected within bright extended sources, especially SNRs, though they
pass the above criteria. About 2\% of the sources are finally excluded
due to problems such as optical loading, serious event pileup, and
being too close to other bright sources.

In the catalog, each source has a unique source number (\verb|SRCID|
in the catalog). However, from visual inspection, we still found 42
sources assigned with multiple SRCID numbers
(Table~\ref{tbl:merge-src}). Each source will be referred to using the
first number in Table~\ref{tbl:merge-src}, but its detections consist
of all the detections corresponding to the different SRCID numbers
(Table~\ref{tbl:A1}). In the end, we have 4330 unique sources in our
sample (Table~\ref{tbl:dercat}).

\tabletypesize{\scriptsize}
\setlength{\tabcolsep}{0.02in}
\begin{deluxetable}{l|l|l|l}
\tablecaption{Sources with multiple SRCID numbers\label{tbl:merge-src}}
\tablewidth{0pt}
\startdata
\hline
 106501 241552 &111034 111032 &113853 113854 &127424 127426  \\
\hline
 134832 134833 210662 &13940 195048 &143291 143295 &147800 147801  \\
\hline
 150531 150532 &163538 163539 &183730 183731 &190860 190861  \\
\hline
 198805 42983 42980 &201208 201209 &202064 62202 &20302 196589   \\
\hline
 205954 86092 &206123 206125 &20843 196637  &210401 245122  \\
\hline
 213162 149986 &221841 4783 &227999 198991&235153 73445  \\
\hline
 235311 73852 &252232 156258 &37588 37584 &49466 49454  \\
\hline
 50074 50083 &50260 50265 &50339 50335 &53515 53514  \\
\hline
 54078 54077 &61618 61621 &6320 6324 &63368 63365  \\
\hline
 66168 66171 &69639 69640 &73642 73647 &8091 8087  \\
\hline
 91437 206280 206281 &93326 93329  
\enddata 
\tablecomments{Each cell lists the SRCID numbers that should correspond to the same source.}
\end{deluxetable}

\subsection{Calculation of the Flux and Hardness Ratio}
\label{sec:refhr}
The 2XMMi-DR3 catalog provides the count rates and fluxes of each
detection for various combinations (individual and total
bands/cameras). The count rates were corrected for various
instrumental effects including the mirror vignetting, detector
efficiency and point spread function losses. The fluxes in individual
energy bands were calculated by dividing the count rates by the energy
conversion factors (ECFs). The ECFs depend on the camera, the filter,
and the source spectrum, which the catalog assumed to be an absorbed
power-law (PL) with a photon index of $\Gamma_{\rm PL}=1.7$ and an
absorption column density of $N_{\rm H}=3\times 10^{20}$ cm$^{-2}$
\citep{wascfy2009}. The ECFs used in the catalog were based on the
calibration before 2008, except for a few observations. We applied
correction factors provided on the catalog website to obtain fluxes
corresponding to the newer calibration. In this paper, we mostly used
the EPIC fluxes in 0.2--12.0 keV (band 8) and 0.2--4.5 keV (called
band 14 hereafter). They were obtained by first summing the fluxes in
the relevant individual bands and then averaging over the available
cameras weighted by the errors. We estimated their systematic errors
to be 0.107 and 0.074, respectively (Appendix~\ref{sec:fhrerr}).

We also calculated four X-ray hardness ratios HR1--HR4 defined as
\begin{equation}
{\rm HR}i=(R_{i+1}-R_i)/(R_{i+1}+R_i),\label{eq:hr}
\end{equation}
where $R_i$ and $R_{i+1}$ are the MOS1-medium-filter equivalent count
rates in the energy bands $i$ and $i+1$, respectively. These count
rates were obtained by multiplying the EPIC fluxes by the ECFs for the
MOS1 camera with the medium filter. We note that the 2XMMi-DR3 catalog
also provides the camera-specific X-ray hardness ratios, but the count
rates used, and thus the hardness ratios, depend on the camera and the
filter. This poses a problem to use the hardness ratios to investigate
the source spectral shape uniformly, considering that the observations
used to create the catalog have various filters and available
cameras. Our hardness ratios are less subject to this problem, as the
fluxes have been calculated to mitigate their dependence on the camera
and filter by using the camera- and filter-dependent ECFs. We only
observed small systematic differences between the pn and MOS hardness
ratios using our definition (Appendix~\ref{sec:fhrerr}).

\subsection{Variability Calculation}
The variability properties can provide important clues about the
source type. There are many kinds of variability. For the short-term
variability that can be measured within a single observation, we paid
attention to stellar X-ray flares, which are good indicators of
coronally active stars \citep{gu2004}. They show a variety of
profiles, though most of them have a fast rise and a slow decay and
can last from minutes to hours. The light curves created for bright
detections by the pipeline are suitable for stellar X-ray flare
search. They are extracted from a circular region of radius
$28\arcsec$ and have bin sizes an integer multiple of 10 s (the
minimum is 10 s), depending on the source intensity
\citep{wascfy2009}. We describe our search for stellar X-ray flares in
Appendix~\ref{sec:flaresearch}.

We also measured the long-term flux variability over different
observations. Both bands 14 and 8 were used, but we will explore
results with band 14 in more detail because the flux in band 5
(4.5--12.0 keV) tends to have large uncertainties. We defined the
long-term flux variability as $V_{\rm var}=F_{\rm max}/F_{\rm min}$
and the significance of the difference as
\begin{equation}
S_{\rm var}=\frac{F_{\rm max}-F_{\rm min}}{(\sigma_{\rm max}^2+\sigma_{\rm min}^2+(rF_{\rm max})^2+(rF_{\rm min})^2)^{1/2}},\label{eq:svar}
\end{equation}
where $F_{\rm max}$ and $F_{\rm min}$ are the maximum and minimum EPIC
fluxes of a unique source, with the corresponding (statistical) errors
$\sigma_{\rm max}$ and $\sigma_{\rm min}$, respectively, and $r$ is
the systematic error ($r=0.107$ and 0.074 for bands 8 and 14,
respectively; Section~\ref{sec:refhr}). We only used detections with
the flux above four times the error ($\sigma$) when calculating
$F_{\rm max}$, while we used $2\sigma$ as the flux for detections with
the flux less than $2\sigma$ when calculating $F_{\rm min}$.

\subsection{Source Type Identification from the Literature}
\label{sec:srcclslit}
\tabletypesize{\scriptsize}
\setlength{\tabcolsep}{0.08in}
\begin{deluxetable}{llll}
\tablecaption{Source type statistical breakdown\label{tbl:identifiedsrc}}
\tablewidth{0pt}
\tablehead{\colhead{Type} & \colhead{\#} & \colhead{Selection} & \colhead{Main Reference} }
\startdata
\multicolumn{4}{l}{Identified:}\\
\hline
Star & 202 & \nodata & \multirow{5}{*}{SIMBAD, GCVS}\\
OrVr & 252 &\nodata & \\
PrSt & 27 &\nodata &\\
VrSt & 101 & \nodata &\\
FlSt & 29 & \nodata &\\
\hline
Sy1n & 29  & Sp=``S1n'' & \multirow{7}{*}{VV10}\\
Sy1 &  242 & Sp=``S1''\\
Sy2 &  63 & Sp=``S2'', ``S1h'' or ``S1i'' \\
LIN &  12& Sp=``S3'', ``S3b'' or ``S3h'' \\
Bla &  27 & Cl=``B'' \\
QSO\tablenotemark{a} &  250 & Cl=``Q''\\
AGN\tablenotemark{a} & 130 & Cl=``A'' except Sp=``H2''\\
\hline
INS & 7 & Type=``XDINS'' & \multirow{3}{*}{ATNFPC}\\
MGR & 11 & Type=``AXP'' \\
rPsr& 21 & Others\\
\hline
aPsr & 34 & \nodata & \multirow{3}{*}{Literature}\\
Bstr & 18 &\nodata & \\
BHB & 4 & \nodata &\\ 
\hline
CV  & 43 & \nodata & CCB \\
\hline
\\
\multicolumn{4}{l}{Candidate:}\\
\hline
Star & 644 & \nodata & This work \\
\hline 
AGN  & 1292 & \nodata & This work\\
G    & 84 & \nodata & This work \\
\hline
SNR & 16 & \nodata &Literature\\
\hline
Mixed & 19 & \nodata &Literature\\
\hline
ULX & 100 & \nodata &Literature\\
\hline
CO & 202 & \nodata &\multirow{3}{*}{This Work}\\
XGS &320 & \nodata \\
GPS & 151 &\nodata 
\enddata 
\tablecomments{See Sections~\ref{sec:srcclslit} and ~\ref{sec:srcuid} for the detailed description of each source type. }
\tablenotetext{a}{Excluding identified Seyfert galaxies, LINERs, or blazars.}
\end{deluxetable}

We established a subsample of our source list with well confirmed
source types from the literature (Tables~\ref{tbl:identifiedsrc} and
\ref{tbl:dercat}), as described below. We refer to them as identified
sources hereafter. There are three main categories of X-ray sources:
stars, active galactic nuclei (AGN) and compact object systems
containing white dwarfs (WDs), neutron stars (NSs) and stellar-mass
black holes (BHs).

For stars, we used the General Catalog of Variable Stars \citep[GCVS,
  Version 2011 January,][]{sadu2009} and sources from SIMBAD (three
from the literature) with optical spectral types available. We
classified stars into the following categories: Orion variables
(OrVr); pre-main-sequence stars (PrSt); variable stars (VrSt,
excluding Orion variables and flaring stars, mostly eclipsing
binaries, rotationally variable and pulsating variable stars); flaring
stars (FlSt); and stars for the rest, typically main-sequence stars.
 
For AGN, we used the catalog of quasars and active nuclei by
\citet[][VV10 hereafter]{veve2010}, with an additional three from the
literature (based on optical spectra). We classified them into seven
main categories based on the classifications given in this catalog
(Table~\ref{tbl:identifiedsrc}): narrow-line Seyfert 1 (Sy1n); Seyfert
1 (Sy1); Seyfert 2 (Sy2); low-ionization nuclear emission-line regions
(LINERs or LINs); blazars (Bla); quasi-stellar objects (QSOs, i.e.,
quasars); and AGN for the rest. There are 11 Seyfert galaxies with
broad polarized Balmer lines (Sp=``S1h'' in VV10) or broad Paschen
lines in the infrared (Sp=``S1i''). They are generally classified as
Seyfert 2 galaxies in SIMBAD, and we adopt the same scheme.

The Australia Telescope National Facility Pulsar Catalog \citep[ATNFPC, as of 2011
  March,][]{mahote2005}\footnote{http://www.atnf.csiro.au/research/pulsar/psrcat}
was used to identify thermally cooling isolated NSs (INSs),
rotation-powered pulsars (rPsr), and magnetars (MGRs). The
accretion-powered X-ray pulsars (aPsr) and bursters (Bstr, believed to
be weakly magnetized accreting NSs (mostly low-mass X-ray binaries))
were identified from the literature, with the requirements of
detections of pulsations and type-I X-ray bursts, respectively. We
referred to \citet{mcre2006} to identify four BH X-ray binaries
(BHBs), which either have secure BH mass measurements or show X-ray
properties very similar to those of BHBs with secure mass measurements
\citep[grade A in Table~4.3 in][]{mcre2006}. The Catalog of
Cataclysmic Binaries (CCB) from \citet[][7th edition]{riko2003} was
used to identify the cataclysmic variables (CVs). All these types of
sources will be generally referred to as compact objects (COs) in this
paper.

We also found 16 sources that are probably (Galactic or
extragalactic) SNRs (or their knots) from the literature. The 19
``mixed'' sources in Table~\ref{tbl:identifiedsrc} (see also
Table~\ref{tbl:dercat}) include two cooling WDs, three supernova, four
gamma-ray bursts, four symbiotic stars, two micro-quasars, etc. The
100 ultra-luminous X-ray sources (ULXs, with luminosity above
$10^{39}$ erg~s$^{-1}$) are either from the literature or from this
work (Section~\ref{sec:srcuid}). We treated all these SNRs, ``mixed''
objects, and ULXs as candidate (instead of identified) sources,
considering that their X-ray mechanisms are largely still unclear.

\subsection{Multi-wavelength Cross-correlation}

We cross-correlated our source X-ray positions with the USNO-B1.0
Catalog \citep{moleca2003} and the 2MASS Point Source Catalog
\citep[2MASS PSC,][]{cuskva2003} to search for their optical and IR
counterparts, respectively. Our sources have small positional errors,
with 86\% less than 0\farcs5 and 99\% less than 1\farcs0 (statistical
plus systematic). The USNO-B1.0 Catalog and the 2MASS PSC have
comparable positional errors. We selected the counterpart to be the
closest one within 4$\arcsec$. For a few sources, we did not assign
the counterparts in this way. One common case is bright extended
galaxies, for which the USNO-B1.0 Catalog and the 2MASS PSC often have
many entries, and we chose the brightest nearby one (in terms of the
$R2$/$K_{\rm s}$-band magnitude) as the counterpart. The other common
case is stars with large proper motions (indicated in the USNO-B1.0
Catalog). In the end we found optical and IR counterparts for 3014
(70\%) and 2058 (48\%) of our sources, respectively
(Table~\ref{tbl:dercat}). For very few sources, we denoted the match
quality as bad due to their coincidence with the persistence artifact
positions in the
2MASS\footnote{http://www.ipac.caltech.edu/2mass/releases/allsky/doc/sec4\_7.html},
bright globular clusters, etc. and we will not explore their
optical/IR properties.

We used the $R2$-band magnitude to define the optical flux as
$\log(F_{\rm O})=-R2/2.5-5.37$, following \citet{magiwo1988}. In some
cases when the $R2$-band magnitude is not available, we used the
$R1$-band magnitude instead. For sources without optical counterparts,
we assumed $R2=21.0$, which is about the minimum among the
counterparts found and was taken as an upper limit. The IR flux was
calculated with the $K_{\rm s}$-band magnitude from the 2MASS PSC:
$\log(F_{\rm IR})=-K_{\rm s}/2.5-6.95$, assuming the ``in-band''
zero-magnitude flux of the $K_{\rm s}$ band from \citet{cowhme2003}.
For sources without IR counterparts, we calculated the upper limit
using $K_{\rm s}=15.3$, the 3-$\sigma$ limiting sensitivity of this
band in the 2MASS. Throughout the paper, all fluxes are in units of
erg~s$^{-1}$~cm$^{-2}$ and correspond to apparent/absorbed values.

We also similarly cross-correlated our sources with the extended IR
sources from the 2MASS Extended Source Catalog \citep[2MASS
  XSC,][]{skcust2006} and the optical sources from the Sloan Digital
Sky Survey \citep[SDSS,][]{abadag2009}. We used them mostly to check
whether our sources are coincident with the center of (extended)
galaxies. To reduce the contamination, we only assumed extended IR
sources from the 2MASS XSC to be galaxies when they are coincident
with the center of an optical extended source, based on visual
inspection of the Digital Sky Survey (DSS) images. For the SDSS
extended sources, we assumed them to be galaxies if their $r$-band
magnitudes are $>$20 (the separation between the point and extended
sources by the SDSS pipeline is not reliable for very faint sources
\citep[see][]{scjodo2002}).

\subsection{Association with Nearby Galaxies}
We used the Third Reference Catalog (RC3) of galaxies
\citep{dedeco1991} to investigate the probability of our sources being
non-nuclear extragalactic sources. This catalog is reasonably
complete for galaxies having apparent diameters larger than 1$\arcmin$
at the $D_{25}$ isophotal level and total $B$-band magnitudes brighter
than about 15.5 and redshifts less than 0.05. The $D_{25}$ isophote
refers to the 25 mag arcsec$^{-2}$ $B$-band isophote and is
approximated as an ellipse. This catalog also includes some objects of
special interest, resulting in 23,022 galaxies in total. It gives the
apparent major and minor diameters and the position angle of the
$D_{25}$ isophote for each galaxy. The $D_{25}$ isophote is roughly
the domain of a galaxy as seen in the DSS images, and by comparing the
source positions with this isophote we can check whether the sources
are possibly associated with that galaxy. Following \citet{libr2005},
we calculated the angular separation $\alpha$ between the galaxy
center and the source, which was then scaled by the elliptical radius
$R_{\rm 25}$ of the $D_{25}$ isophotal ellipse in the direction from
the galaxy center to the source. We assumed that only sources with
$\alpha/R_{\rm 25}$$<$2 and $\alpha$$>$4$\arcsec$ can possibly be
non-nuclear extragalactic sources. The positions of the galaxy
centers given in the RC3 have only modest accuracy. We found that the
positions from NASA/IPAC Extragalactic Database
(NED)\footnote{http://ned.ipac.caltech.edu} are closer to the real
galaxy centers (at least for galaxies related to our sources) based on
visual inspection of DSS images. Thus we chose to use the galaxy
center positions from NED when possible. We also estimated the maximum
0.2--12.0 keV absorbed luminosity for each candidate non-nuclear
extragalactic source. We adopted the distances to most galaxies used
in \citet{libr2005} and \citet{li2011}. For 17 other galaxies, we
obtained their distances using the redsifts from NED and assuming a
flat universe with the Hubble constant $H_0$=75 km s$^{-1}$ Mpc$^{-1}$
and the matter density $\Omega_{\rm M}$=0.27 if they have recessional
velocities larger than 1000 km~s$^{-1}$ or from the literature
otherwise.

\subsection{Simple Spectral Fits}
\label{sec:smspfit}
We created simple energy spectra, one for each available camera for
each detection, using the count rates in five basic energy bands in
the 2XMMi-DR3 catalog and fitted them with an absorbed PL
(Table~\ref{tbl:A1}). The response files corresponding to the
``thin'', ``medium'', and ``thick'' optical blocking filters were
constructed using the observations 0112560101, 0204870101, and
0311190101, respectively, from source regions centering at the
pointing direction and covering the whole field of view. SAS 11.0.0
and the calibration files of 2011 June were used. The event selection
criteria for each energy band/camera used for source detection in
\citet{wascfy2009} were followed. The interstellar medium absorption
was modeled by the WABS model in XSPEC. A systematic error of 10\% on
the model was assumed. Relative normalizations among different cameras
were allowed to be free. We note that we used the PL model just to
roughly characterize the spectral shape. The real detailed X-ray
spectra of AGN are still often described by this model when there are
no strong soft excesses, but for stars, the APEC and MEKAL models in
XSPEC are generally needed to describe their detailed X-ray spectra
well.

\begin{figure*}
\includegraphics[width=0.99\textwidth]{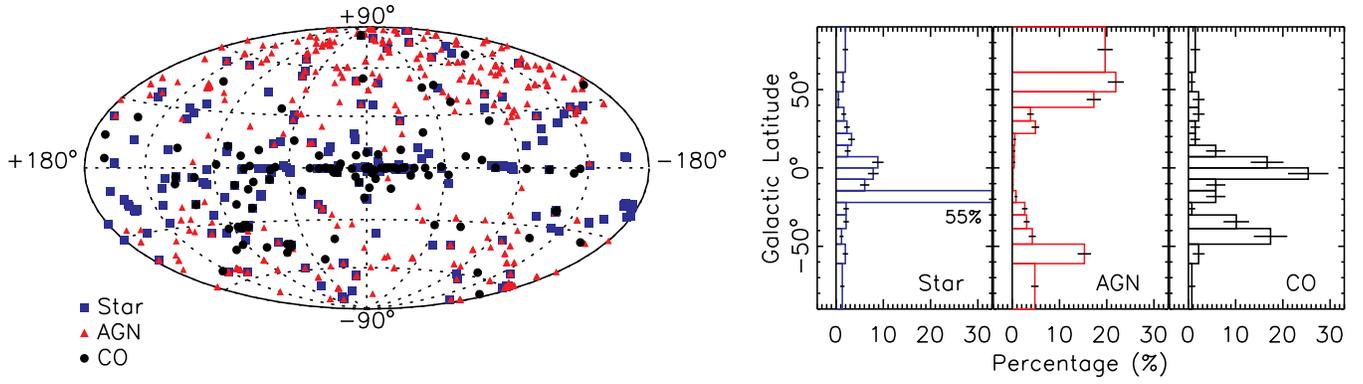}
\caption{Hammer-Aitoff equal area projection in Galactic coordinates of identified sources (left) and their distributions with respect to the Galactic latitude (right).  \label{fig:xraycoormap}}
\end{figure*}

\begin{figure*}
\subfigure[Stars]{
\includegraphics[width=0.32\textwidth]{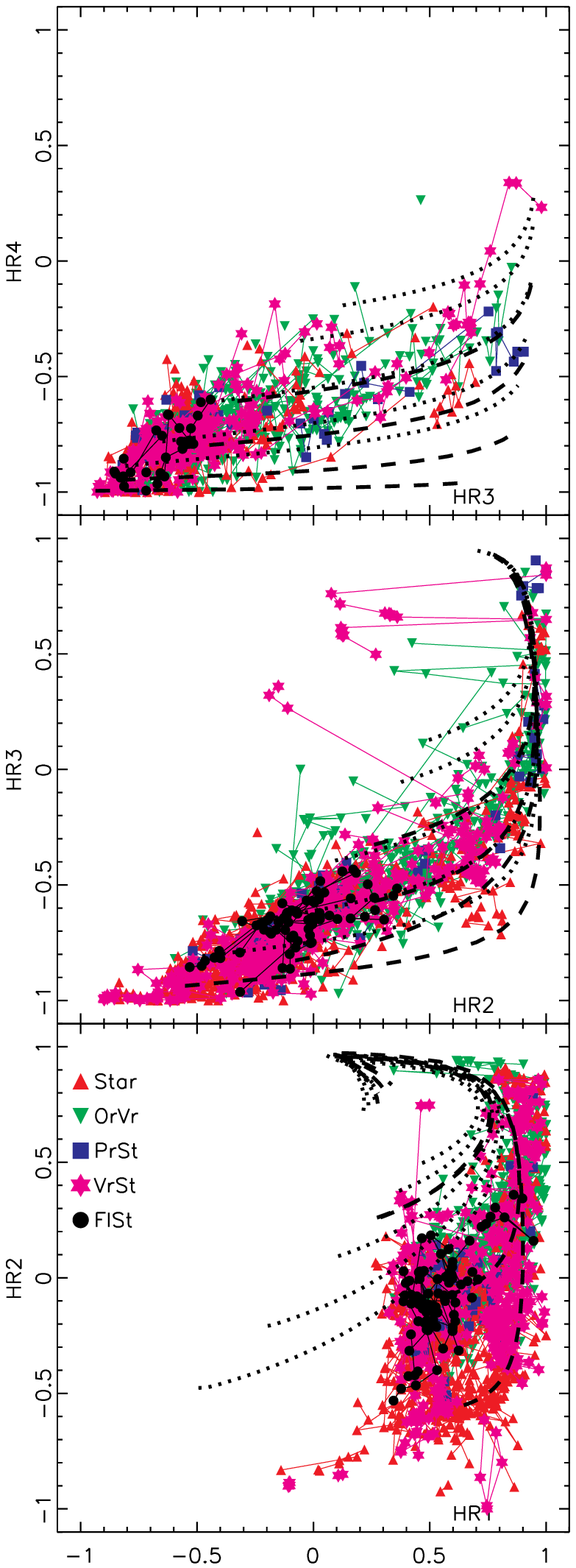}
\label{fig:starcolor}
}
\subfigure[AGN]{
\includegraphics[width=0.32\textwidth]{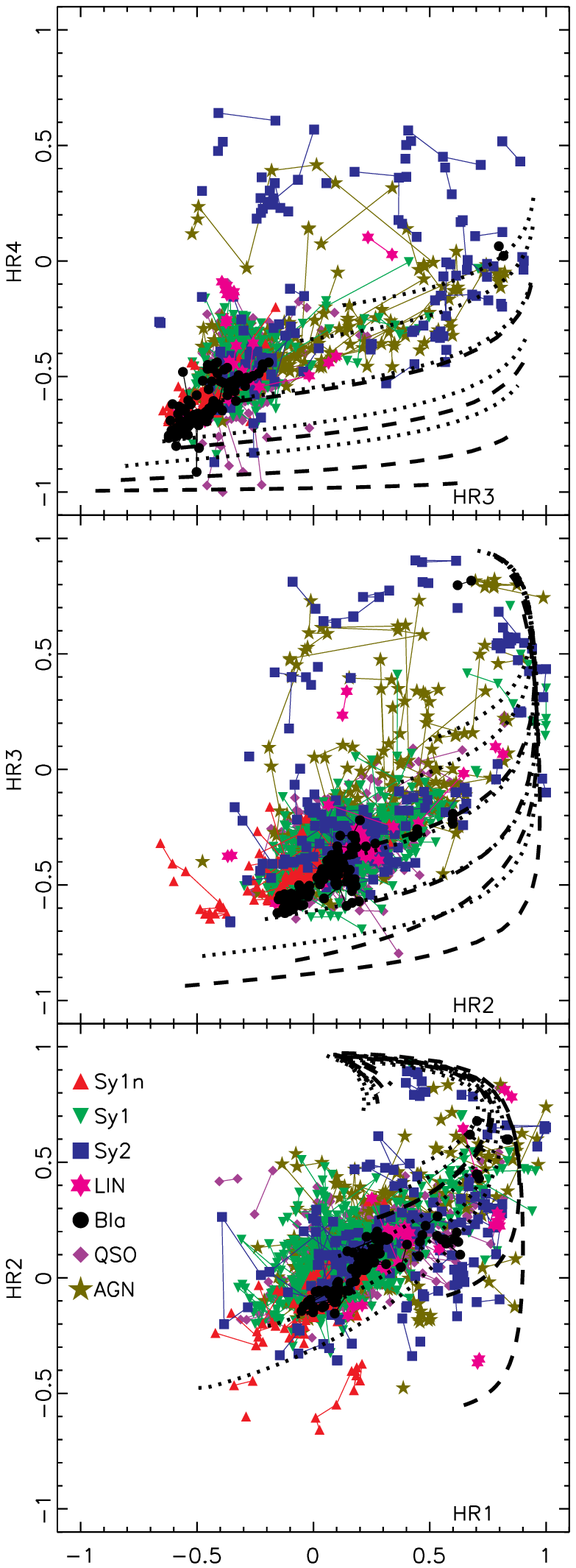}
\label{fig:agncolor}
}
\subfigure[Compact objects]{
\includegraphics[width=0.32\textwidth]{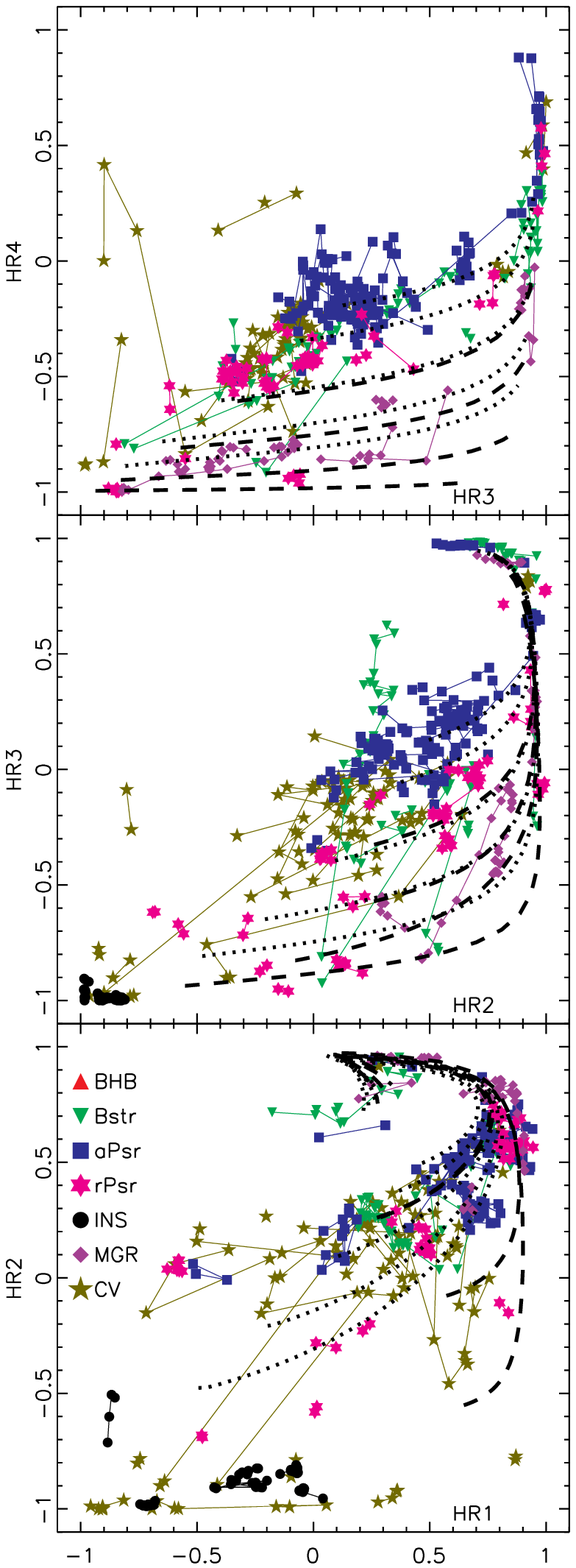}
\label{fig:cocolor}
}
\caption{The X-ray color-color diagrams for identified sources. Overplotted are PL spectra (dotted lines) with $\Gamma_{\rm PL}=0.5$ (top), 1, 2, 3, and 4, and thermal spectra (APEC model with 0.5 solar abundance, dashed lines) with temperatures of 0.5 (bottom), 1, 2, and 5 keV. They are obtained with $N_{\rm H}$ varying from 0 (lower-left) to $10^{23}$ cm$^{-2}$. The detections for each source are connected by solid lines in an increasing order of HR3.  \label{fig:xraycolor}}
\end{figure*}

\section{RESULTS}

\subsection{Properties of Identified Sources}
\label{sec:res}

\subsubsection{Spatial distribution}
\label{sec:spdis}

We show the spatial distribution of the identified sources in Galactic
coordinates in Figure~\ref{fig:xraycoormap} (left panel) and their
distributions with respect to the Galactic latitude $b$ (right
panel). We differentiate stars, AGN, and compact objects. We note that
in the sky map in the left panel some fields are too crowded to see
the real number of sources, while the distributions on the right panel
complement this information. The bin sizes in the distribution plots
vary to correspond to an equal spatial area. The error bars shown were
obtained by assuming Poisson statistics in each bin (this is also
assumed for all other distribution plots throughout this paper). We
see that there are many stars near $b=-19.4\degr$, which is due to the
Orion Nebula (about 42\% of the identified stars are from this
nebula). Apart from this field, the identified stars concentrate at
low Galactic latitudes. Stars with high Galactic latitudes are also
seen (about 31\% with $|b|>20\degr$, excluding those from the Orion
Nebula). Stars are typically in the Galactic plane. The large spread
of stars along the Galactic latitude in our sample indicates that they
are nearby.

The identified AGN tend to be at high Galactic latitudes, only about
2\% with $|b|<20\degr$. Such a bias can be explained by the large
absorption along the Galactic plane and the preference to target the
region outside the Galactic plane in galaxy surveys, such as the
SDSS. There are more identified AGN from the northern Galactic
hemisphere (about 68\% with $b>20\degr$ while about 30\% with
$b<-20\degr$), which can be explained by the fact that many AGN in our
sample are discovered by the SDSS, whose sky access is primarily in
the north.

The identified compact objects clearly concentrate near the low
Galactic latitude region, about 47\% with $|b|<10\degr$. The excess of
the identified compact objects in the region of
$-50\degr<|b|<-30\degr$ is due to the presence of LMC and SMC.

\subsubsection{X-ray color-color diagrams and spectral fits}
\label{sec:xcdsf}

Figure~\ref{fig:xraycolor} shows the X-ray color-color diagrams for
identified sources. Stars, AGN, and compact objects are plotted in the
left, middle, and right columns, respectively. Only data points with
error bars less than 0.1 on both colors are plotted. Spectral tracks
of absorbed PL (dotted lines) and APEC spectra (dashed lines) are
overlaid for comparison. APEC is a model of the emission spectrum from
collisionally-ionized diffuse gas in XSPEC and is often used to model
the X-ray spectra of stars. In the HR1-HR2 diagram, we see that most
stars occupy a region with characteristic APEC temperatures about
0.5--1 keV. However, in the HR2-HR3 and HR3-HR4 diagrams, there are
many stars in a region with APEC temperatures $>$1 keV, which means
that these stars probably need multiple APEC components to describe
them. Investigations of these sources show that a majority of them
correspond to detections with flares. Stars rarely show very hard
X-ray spectra. Be stars such as HD 110432 (\#110586) and SS 397
(\#167002) can be very hard \citep[e.g.,][]{lomosm2007}, with HR4
larger than $-0.4$ in Figure~\ref{fig:xraycolor}. The outliers in the
upper-left region in the HR3-HR2 diagram consist of famous stellar
systems such as $\eta$ Carinae (\#87820), $\gamma^2$ Velorum
(\#65891), HL TAU (\#38943), EX Lupi (\#248785) and FU Orionis
(\#53697)
\citep[e.g.,][]{fawale2011,scgume2004,gifasi2006,grhaka2010,skbrgu2006}.

Figure~\ref{fig:xraycolor} shows that AGN (middle column) are mostly
in the region with $\Gamma_{\rm PL}=$1--3.  The main outliers are
Seyfert 2, which have HR4 larger than 0. They may have strong
absorption and typically have two main components with the hard one
probably due to the reflection and/or heavy absorption of Seyfert 1
nuclei \citep{tugena1997,awmuog2006,noteaw2009}. Comparing the HR1-HR2
and HR3-HR4 diagrams, we see that most narrow-line Seyfert 1 galaxies
are in a region with $\Gamma_{\rm PL}>$2 in the HR1-HR2 diagram, but
in a region with $\Gamma_{\rm PL}<$2 in the HR3-HR4
diagram. This can be explained by the frequent presence of soft
excesses in these sources.

Compact objects show a large spread in Figure~\ref{fig:xraycolor}
(right column). This is to a large degree due to many subclasses
included. However, many types of compact objects are known to show
large spectral variations. For example, some CVs (brown stars) can
have spectra changing from very hard to super-soft. Some weakly
magnetized accreting NSs (bursters) also have a large range of HR3
(green triangles). Accretion-powered X-ray pulsars, whose X-ray
spectra are known to be hard \citep{whswho1983}, occupy the region
with $\Gamma_{\rm PL}$ around 0.5, where we see few stars and AGN. The
thermally cooling isolated NSs (black filled circles) stay in the
lower left corner of the (HR1-HR2 and HR2-HR3) diagrams, as expected
due to their soft nature \citep{ha2007}. The magnetars (purple
diamonds) look very hard in the HR1-HR2 diagram, which can be
explained by large absorption, but many of them occupy a soft region
in the HR3-HR4 diagram, with $\Gamma_{\rm PL}\gtrsim$3. Many magnetars
are known to show strong spectral turnover at 15 keV, with very soft
and very hard spectra below and above this energy, respectively
\citep{ennama2010,kuhede2006}, and our results are consistent with
this.

\begin{figure}
\centering
\includegraphics[width=0.45\textwidth]{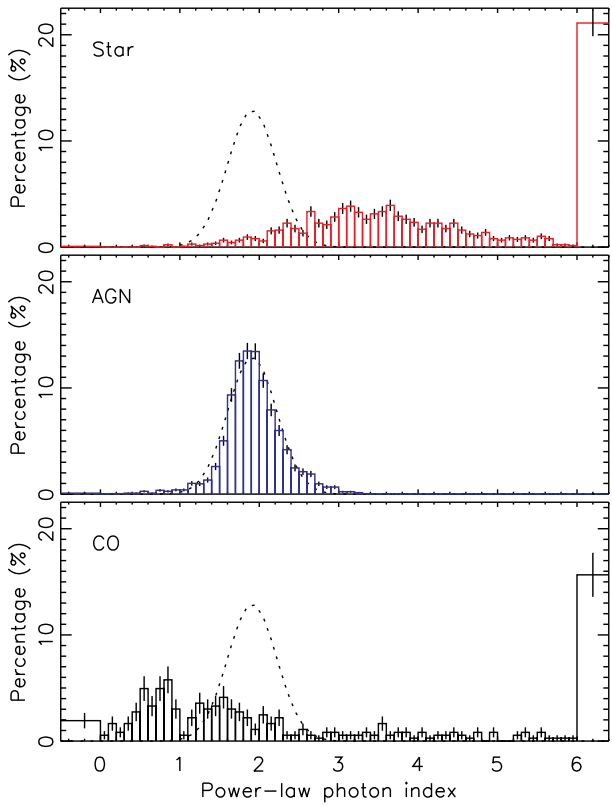}
\caption{The $\Gamma_{\rm PL}$ distribution of identified sources. A normal distribution (dotted line) with a mean of 1.91 and a standard deviation of 0.31 is also plotted in each panel for reference; it is an approximate distribution of $\Gamma_{\rm PL}$ of AGN in the center panel. We accummulate all detections with $\Gamma_{\rm PL}>6$ into one bin and those with $\Gamma_{\rm PL}<0$ into another bin. \label{fig:plfitagn}}
\end{figure}

The spectral differences among various classes of sources can also be
seen from the simple fits with an absorbed PL to spectra created using
the band count rates in the 2XMMi-DR3 catalog
(Table~\ref{tbl:A1}). Focusing on detections with S/N $\ge 14$, we
obtained 1374 (79\%), 2271 (92\%) and 364 (65\%) detections with
reduced $\chi^2$ $\le 4$ for stars, AGN, and compact objects,
respectively. Their $\Gamma_{\rm PL}$ distributions are shown in
Figure~\ref{fig:plfitagn}, and we see that they are remarkably
different for different classes. AGN have $\Gamma_{\rm PL}$
concentrated near the median of 1.91, with a standard deviation of
only 0.31, while stars and compact objects have $\Gamma_{\rm PL}$ much
more scattered. The $\Gamma_{\rm PL}$ values for stars are generally
high, with only about 13\% less than 2.5. We found that a majority of
detections with low $\Gamma_{\rm PL}$ are due to flares, while
detections with $\Gamma_{\rm PL}>6.0$ (corresponding to low APEC
temperatures in the color-color diagrams) are mostly (about 70\%) from
main sequence stars. For compact objects, detections with $\Gamma_{\rm
  PL}>6.0$ are mostly from thermally cooling isolated NSs and novae (a
subclass of CV). There are 31\% of detections with $\Gamma_{\rm
  PL}<1.0$, mostly from accretion-powered X-ray pulsars. In
comparison, AGN and stars have only 1.7\% and 0.5\% of detections with
$\Gamma_{\rm PL}<1.0$, respectively.

\begin{figure*}
\subfigure[Stars]{
\includegraphics[width=0.32\textwidth]{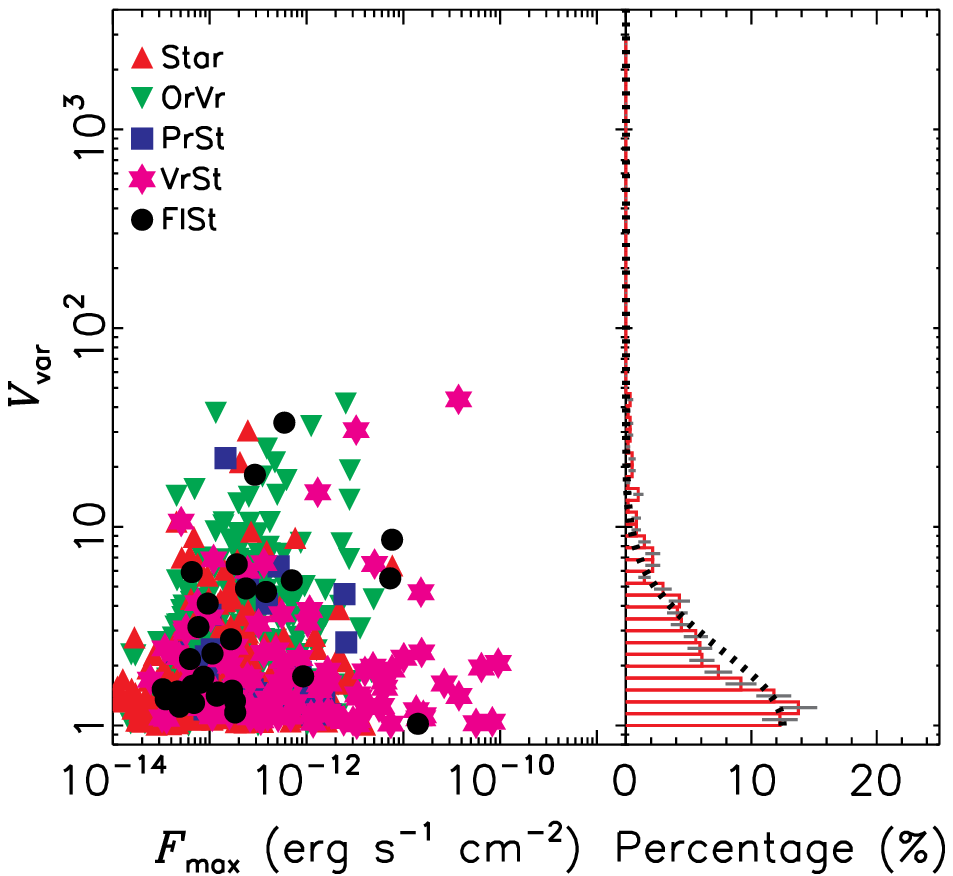}
\label{fig:starfvar}
}
\subfigure[AGN]{
\includegraphics[width=0.32\textwidth]{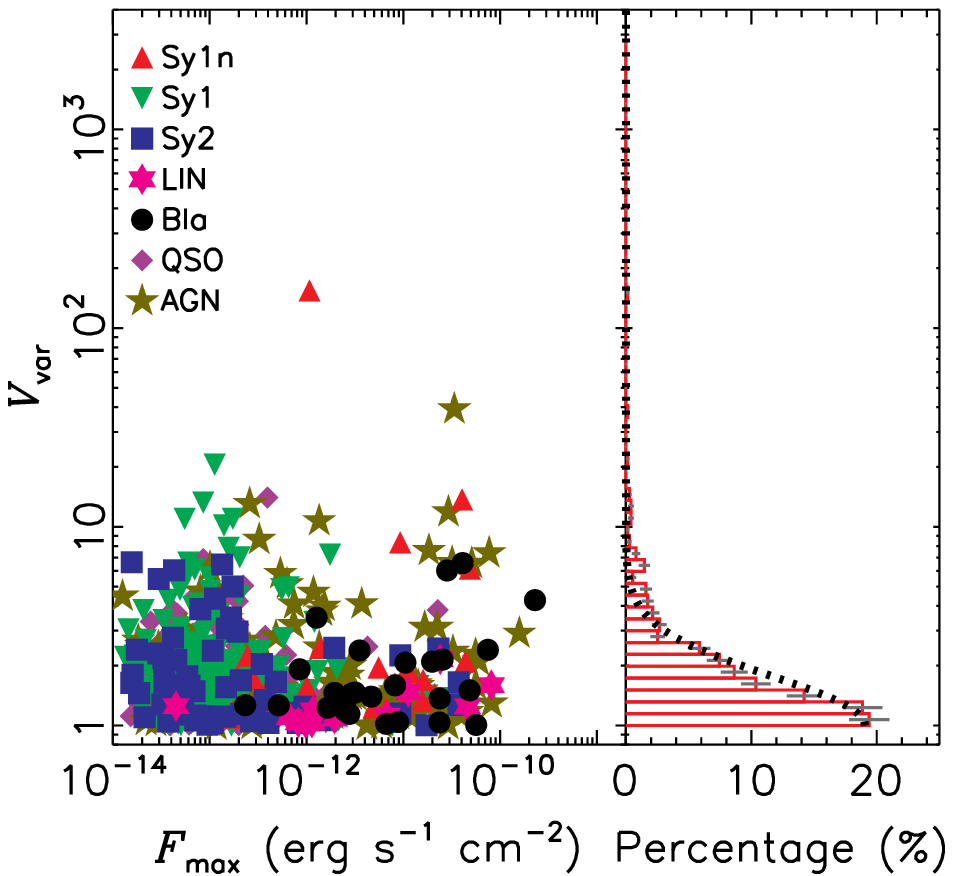}
\label{fig:agnfvar}
}
\subfigure[Compact objects]{
\includegraphics[width=0.32\textwidth]{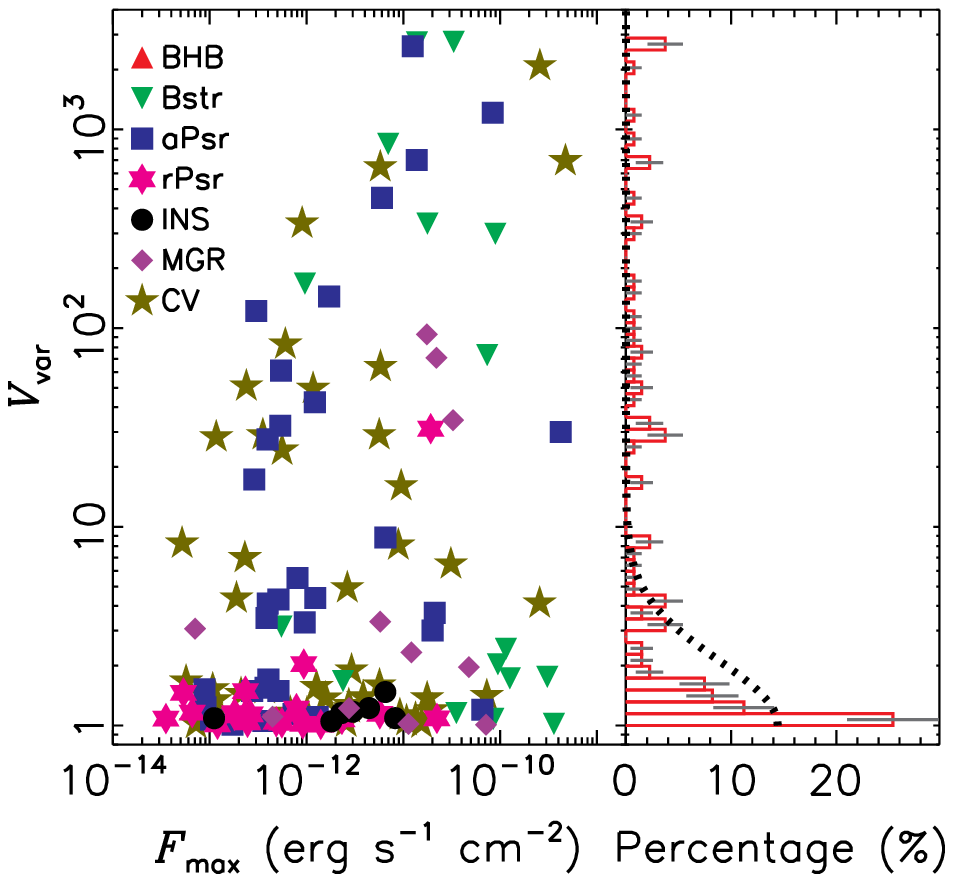}
\label{fig:cofvar}
}
\caption{The X-ray flux variation factor versus the maximum flux in 0.2--4.5 keV and the distribution of the flux variation factors. Two bursters, \#151436 ($V_{\rm var}$=$1.8\times10^4$) and \#163131 ($V_{\rm var}$=$3.0\times10^4$), are outside the plotting range in (c). Half-normal distributions of the flux variation factor logarithms using the same median as the data are overplotted (dotted lines). \label{fig:xrayfvar}}
\end{figure*}

\begin{figure} 
\centering
\includegraphics[width=0.4\textwidth]{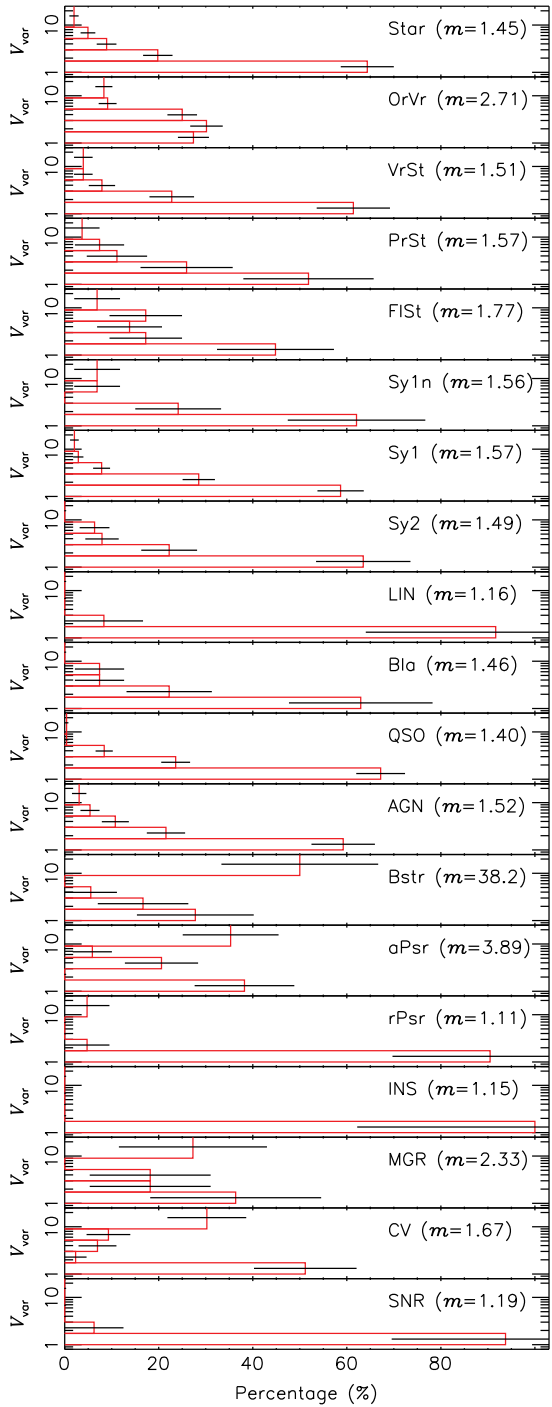}
\caption{The distribution of the X-ray flux variation factors for each source type, with the median value also given in parentheses. The last bin in each panel includes all sources with variation factors $>$9. \label{fig:xrayfvarind}}
\end{figure}

\begin{figure} 
\centering
\includegraphics{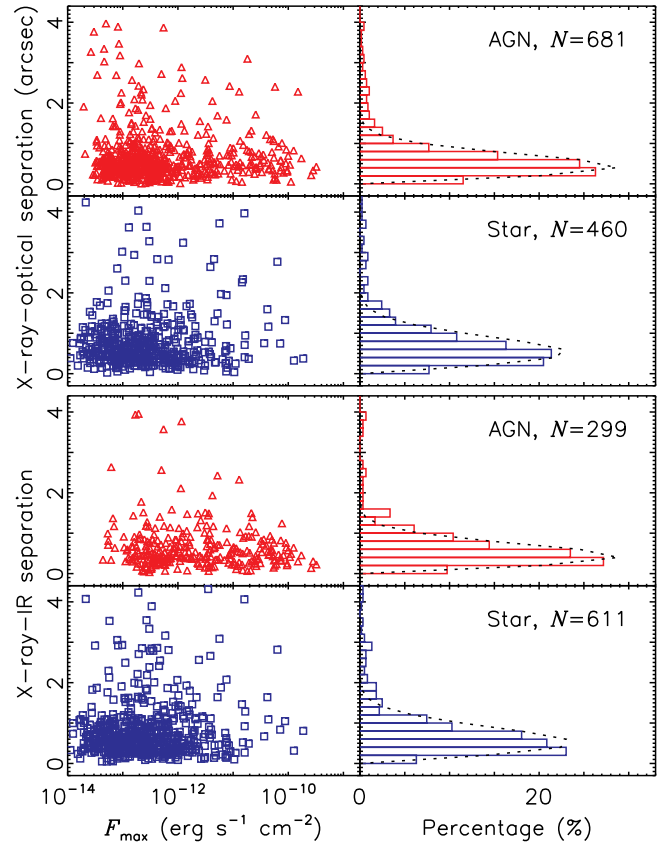}
\caption{The X-ray-optical and X-ray-IR separations versus the maximum 0.2--12.0 keV flux (left panels) and the distributions of the separations (right panels), which are approximated as Rayleigh distributions (dotted lines; see text). \label{fig:xrayoptiroffset}}
\end{figure}

\begin{figure*} 
\centering
\includegraphics{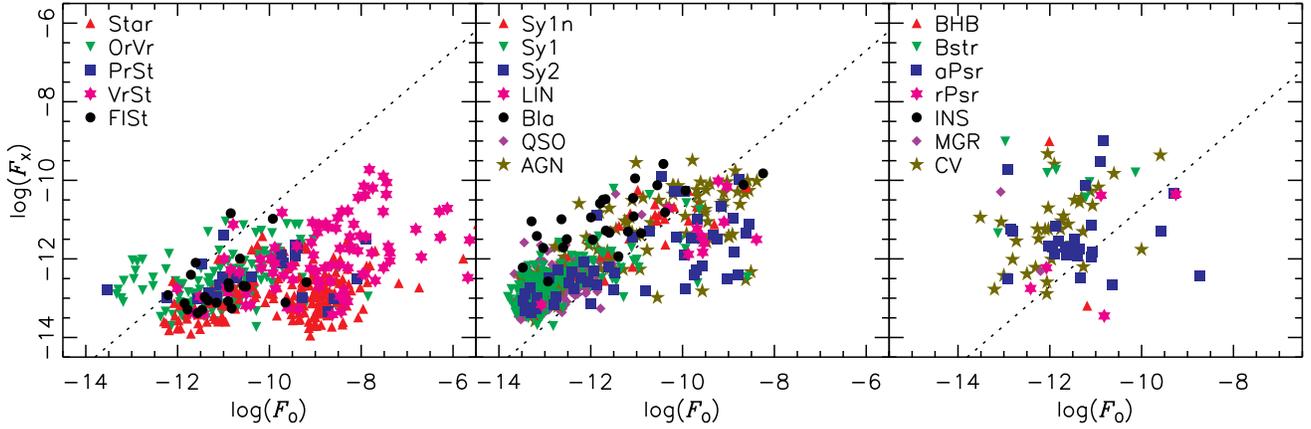}
\caption{The X-ray (0.2--12.0 keV; maximum) versus optical (USNO-B1.0 $R2$-band) fluxes of identified sources with optical counterparts. The dotted reference lines, the same for all panels, are plotted to indicate the separation of stars from other source types. \label{fig:xrayofluxind}}
\end{figure*}

\begin{figure*} 
\centering
\includegraphics{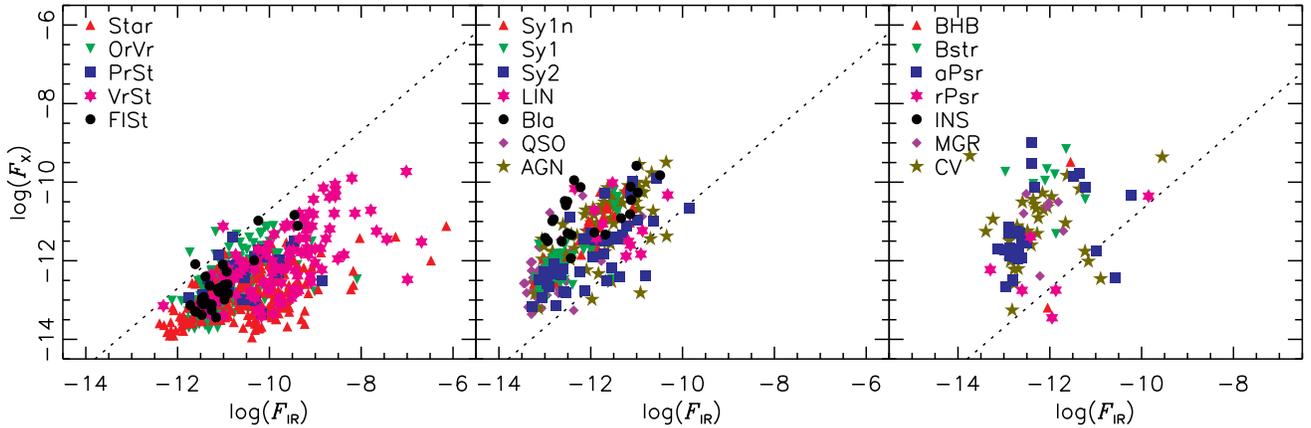}
\caption{The X-ray (0.2--12.0 keV; maximum) versus IR (2MASS $K_{\rm s}$-band) fluxes of identified sources with IR counterparts. The dotted reference lines, the same for all panels, are plotted to indicate the separation of stars from other source types.\label{fig:xrayirfluxind}}
\end{figure*}

\begin{figure*} 
\centering
\includegraphics{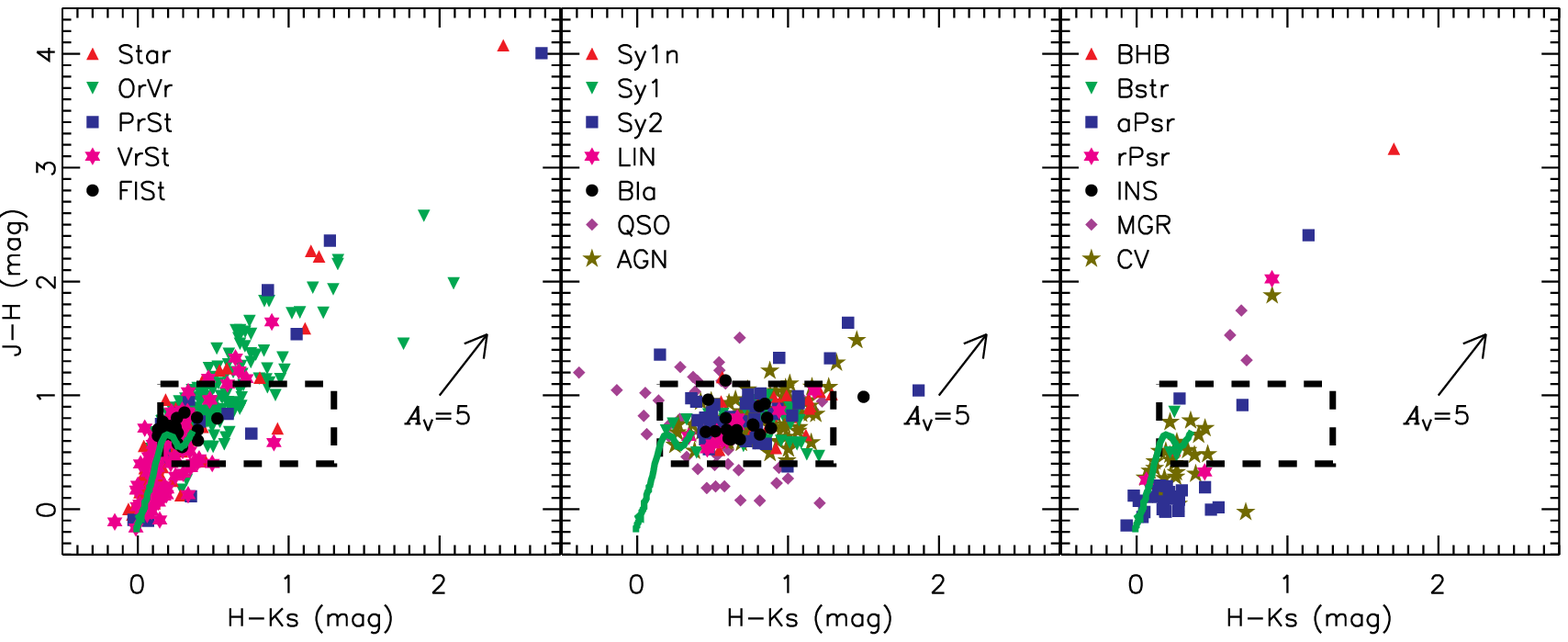}
\caption{The 2MASS color-color diagram of identified sources. Only the counterparts detected in all three 2MASS bands are plotted. One Orion variable (\#169259, with $J-H=4.84$ and $H-K_{\rm s}=3.51$) is outside the plotting range. The big dashed square, which encloses the data points of most AGN, is repeated in all panels for reference. The green solid line represents the locus of main-sequence stars, obtained from \citet{ko1983} after color transformation (http://www.ipac.caltech.edu/2mass/releases/allsky/doc/sec6\_4b.html). Reddening by $A_{\rm V}$=5 \citep{rile1985} is indicated by an arrow. \label{fig:2masscolorind}}
\end{figure*}

\subsubsection{Long-term X-ray variability}
\label{sec:ltxvarid}

Figure~\ref{fig:xrayfvar} plots the X-ray flux variation factor
$V_{\rm var}$ versus the maximum flux in 0.2--4.5 keV and the
distributions of $V_{\rm var}$ for the three main classes (i.e.,
stars, AGN, and compact objects). We obtained median variation factors
$m=1.79$, 1.48, and 1.65 for stars, AGN, and compact objects,
respectively. There are about 95.4\% of the stars, 98.4\% of the AGN,
and 71.7\% of the compact objects with $V_{\rm var}$ less than 10. The
distributions of $V_{\rm var}$ for stars and AGN are relatively
smooth. They roughly resemble half-normal distributions for the
logarithm of $V_{\rm var}$ (dotted lines, with the same median as the
data). When the flux logarithm for each source follows the same normal
distribution and there are only two detections for each source, the
logarithm of $V_{\rm var}$ is expected to follow a half-normal
distribution. In contrast, the distribution of $V_{\rm var}$ for
compact objects shows a large spread and depends on the subclasses
clearly, as discussed below.

The distribution of $V_{\rm var}$ for each subclass is given in
Figure~\ref{fig:xrayfvarind}. We see that Orion variables are more
variable (with $m=2.71$) than other stars (such as the main sequence
stars, which have $m=1.45$). The variable stars, which are classified
mostly due to their optical variability, have $m=1.51$, close to that
of main sequence stars. The three subclasses of Seyfert galaxies
(Sy1n, Sy1, and Sy2) overall show similar variability, with $m=1.56$,
1.57, and 1.49, respectively. However, some narrow-line Seyfert 1
galaxies show very large variability \citep[a factor of 154 for PHL
  1092; see also][]{mifabr2009}. LINERs have $m=1.16$, which is much
lower than those of other AGN, but we cannot exclude the probability
that it is due to contamination of steady circumnuclear diffuse
emission in these systems \citep[e.g.,][]{gomamr2009}.

The large spread of the distribution of $V_{\rm var}$ for compact
objects mostly comes from accreting compact objects (BHBs, bursters,
accretion-powered X-ray pulsars, and CVs;
Figures~\ref{fig:xrayfvar}-\ref{fig:xrayfvarind}). They were observed
to vary by factors up to a few thousands. Some of them are known to
vary much more than measured here (e.g., \object{Aql X-1}). This can
be explained by a selection effect because \xmm\ cannot observe bright
sources in imaging modes. For isolated NSs, the rotation-powered
pulsars and thermally cooling isolated NSs have $m=1.11$ and 1.15,
respectively (their median values of $S_{\rm var}$ are only 0.79 and
1.35, respectively). PSR J1302-6350 (i.e., PSR B1259-63, \#116117) is
the only one with significantly large variability (a factor of
31.2). However, it is a binary system, and its X-ray emission
mechanism is probably different from most rotation-powered pulsars
\citep{chnelu2006}. Different from the above two classes of isolated
NSs, the majority of magnetars are observed to be variable, with some
varying by factors of a few tens.

\subsubsection{Multi-wavelength cross-correlation}
\label{sec:oirmatchid}

We found optical counterparts from the USNO-B1.0 Catalog for about
75\%, 90\%, and 56\% of the identified stars, AGN, compact objects,
respectively. The corresponding percentages of IR counterparts found
from the 2MASS PSC are 100\%, 40\%, and 56\%, respectively. In
Figure~\ref{fig:xrayoptiroffset}, we show the X-ray-optical and
X-ray-IR positional separations for the identified stars and
AGN. About 90\% of the optical/IR counterparts have separations
$<$1\farcs5 from the X-ray positions. Separations are expected to
follow the Rayleigh distribution if both R.A. and Decl. have a
constant separation error $\sigma_{\rm sep}$ (the quadratic sum of the
X-ray and optical/IR positional errors). The value of $\sigma_{\rm
  sep}$ for each Rayleigh distribution overplotted on the distribution
of the separations in Figure~\ref{fig:xrayoptiroffset} (dotted line)
was inferred from the median of the separations, which is 1.18 times
$\sigma_{\rm sep}$ for the Rayleigh distribution. We obtained
$\sigma_{\rm sep}$=0\farcs42 and 0\farcs52 for the optical
counterparts of AGN and stars, respectively. The corresponding values
for the IR counterparts are $\sigma_{\rm sep}$=0\farcs42 and
0\farcs51, respectively. The optical/IR counterparts of stars on the
whole have larger separations from X-ray positions than those of AGN,
probably due to stellar proper motions. The above results indicate a
high astrometric accuracy of our source sample.

The identified compact objects for which we found the optical/IR
counterparts are mostly CVs and accretion-powered X-ray pulsars. We
found no optical/IR counterparts for the seven thermally cooling
isolated NSs. In fact, their optical counterparts are known to be very
faint, with $B$-band magnitudes about 26 \citep{pimoja2009}. Some
optical/IR counterparts are found for magnetars, but they all have
large separations ($>$1\farcs9) from the X-ray positions and are
mostly in the crowded fields of optical/IR sources, thus probably
spurious. Indeed, magnetars generally have $R$-band magnitudes larger
than 24 and $K_{\rm s}$-band magnitudes around 19 \citep{woth2006},
below the detection limits of the USNO-B1.0 Catalog and the 2MASS
PSC. We found few optical/IR counterparts for the 21 rotation-powered
pulsars in our sample either. PSR J1302-6350 (i.e., PSR B1259-63,
\#116117) and PSR J1023+0038 (\#83189) are the only two confident
cases with optical and IR counterparts, but both of them are binary
systems \citep{jomaly1992,arstra2009}.

Figure~\ref{fig:xrayofluxind} plots the X-ray flux versus the optical
flux for identified sources with optical counterparts. It shows that
the identified stars span higher optical fluxes than AGN and compact
objects, but they all have the lowest optical fluxes around
$10^{-13.5}$ erg~s$^{-1}$~cm$^{-2}$, about the detection limit in the
USNO-B1.0 Catalog. Figure~\ref{fig:xrayirfluxind} plots the X-ray flux
versus the IR flux for identified sources with IR counterparts. We see
that the identified stars also span higher IR fluxes than AGN and
compact objects. Moreover, they all have IR fluxes
$\gtrsim$$10^{-12.5}$ erg~s$^{-1}$~cm$^{-2}$, brighter than the
3$\sigma$ detection limit, while many identified AGN and compact
objects have IR fluxes lower than this limit.

In the IR color-color diagram in Figure~\ref{fig:2masscolorind} for
sources with IR counterparts, we see that the identified AGN
concentrate within the big dashed square, which is in agreement with
previous studies \citep[e.g.,][]{hyal1982}. The IR counterparts of the
identified stars and compact objects span a larger range of
colors. Compared with the locus of main-sequence stars (the green
solid line in Figure~\ref{fig:2masscolorind}) and the reddening vector
(the arrow), their large spread in IR colors is probably due to strong
extinction in some of them.
 
\begin{figure} 
\centering
\includegraphics{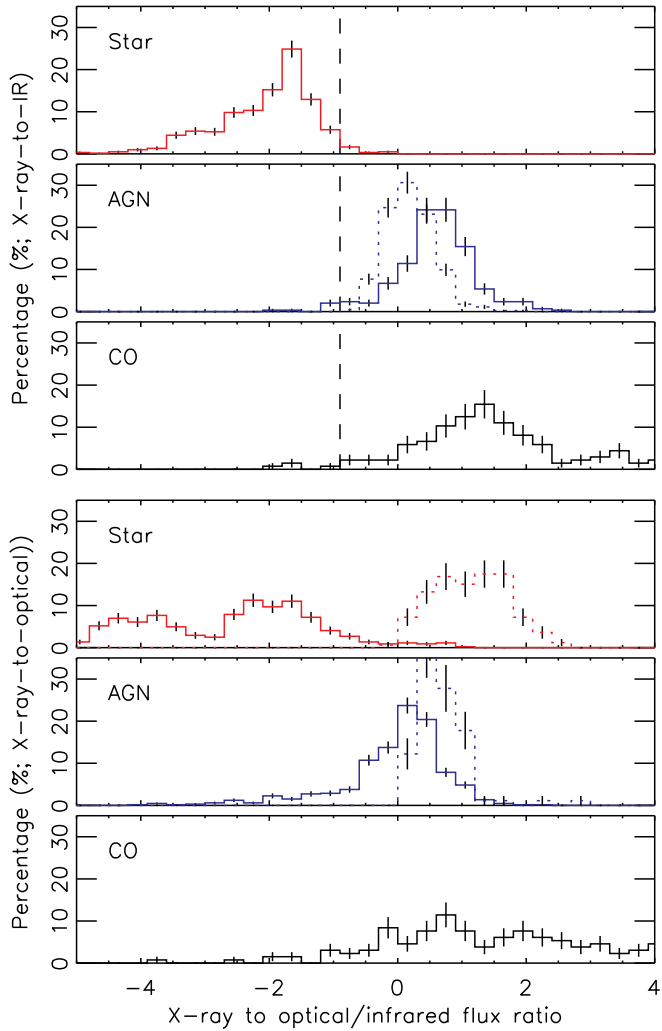}
\caption{The distributions of the X-ray-to-optical (top three panels) and X-ray-to-IR flux ratios (bottom three panels) of identified sources. For stars and AGN, we
separate the sources with (solid line) and without (dotted line) optical/IR counterparts, but not for compact objects. The dashed lines in the distribution plots of X-ray-to-IR flux ratios are for reference.\label{fig:xrayoirhistind}}
\end{figure}

Figure~\ref{fig:xrayoirind} plots the X-ray-to-IR versus X-ray-to-optical
flux ratios for all identified sources. For sources without optical/IR
counterparts, their optical/IR flux upper limits are used and they are
circled in this figure. Sources with neither optical nor IR
counterparts fall onto the dotted reference line. The corresponding
distributions of the X-ray-to-optical and X-ray-to-IR flux ratios are
shown in Figure~\ref{fig:xrayoirhistind}. For stars and AGN, we
separate the sources with (solid line) and without (dotted line)
optical/IR counterparts (there is no distribution plot for stars
without IR counterparts because they all have IR counterparts). As
there are not many compact objects, they are not separated but
combined to create a single distribution plot.  In
Figures~\ref{fig:xrayoirind}--\ref{fig:xrayoirhistind}, we see that
stars are generally separated from AGN and compact objects in terms of
the X-ray-to-IR flux ratio (refer to the dashed reference lines; see
also Figure~\ref{fig:xrayirfluxind}). Stars have $\log(F_{\rm
  X}/F_{\rm IR})$$\lesssim$$-0.9$, AGN $-0.9$$\lesssim$$\log(F_{\rm
  X}/F_{\rm IR})$$\lesssim$2.5, and compact objects $\log(F_{\rm
  X}/F_{\rm IR})$$\gtrsim$0.5. AGN tend to have higher
X-ray-to-optical flux ratios ($\log(F_{\rm X}/F_{\rm
  O})$$\gtrsim$$-1.0$) than stars ($\log(F_{\rm X}/F_{\rm
  O})$$\lesssim$$-1.0$) for sources with optical
counterparts. However, many stars, especially Orion variables, have no
optical counterparts. These stars are probably subject to strong
extinction so that their optical counterparts fall below the detection
limit of the USNO-B1.0 Catalog, making their (apparent)
X-ray-to-optical flux ratios as large as those of AGN and compact
objects.

Figure~\ref{fig:xrayoirhistind} shows that, compared with AGN with
optical counterparts, AGN without optical counterparts on the whole
have slightly larger X-ray-to-optical flux ratios, which were obtained
using an (assumed) optical upper limit of $R2=21.0$. This is probably
due to the selection bias that AGN without optical counterparts are
far away and tend to be included by us if they have high
X-ray-to-optical flux ratios and are thus bright in X rays, and/or the
selection bias that AGN with optical counterparts include sources with
strong star-forming activity in the host galaxies and thus with large
star light contamination.

\subsection{The Source Type Classification Method}
\label{sec:clsscrit}

We were able to identify the source types from the literature for only
about one third of our sources. For the rest, we classified them as
one of three candidate classes: stars, AGN, and compact objects. We
derived the following source type classification method.

Sources with $\log(F_{\rm X}/F_{\rm IR})$$<$$-0.9$ are classified as
stars, except those with HR1$\le$0.3 and those in the direction of
centers of galaxies (from the RC3, the 2MASS XSC, the SDSS,
etc.). Sources are treated as in the direction of galactic centers if
the separation is $<$4$\arcsec$. Sources with $\log(F_{\rm X}/F_{\rm
  IR})$$>$$-0.9$ are classified as stars if they have X-ray flares
detected.

The remaining sources are assumed to be either AGN or compact
objects. Sources with HR1$<$$-0.4$, HR2$<$$-0.5$, HR3$<$$-0.7$,
or HR4$<$$-0.8$ (soft sources), sources with $-0.1$$<$HR3$<$0.5 and
$-0.25$$<$HR4$<$0.1 (hard sources like accretion-powered X-ray
pulsars), sources with $V_{\rm var}$(0.2--4.5 keV)$>$10, and sources
with $\log(F_{\rm X}/F_{\rm IR})$$>$2.5 are classified as compact
objects and denoted as strong candidates. Sources with only a small
fraction of detections marginally passing the above X-ray color
selection criteria are not classified as compact objects using the
colors. To decrease the probability of missing compact objects,
remaining sources that are probably non-nuclear extragalactic sources
($\alpha/R_{\rm 25}\le2.0$ and $\alpha> 4\arcsec$) or probably in the
Galactic plane ($|b|\le 10\degr$) are also classified as compact
objects but denoted as weak candidates.

Applying the above classification scheme to the identified sources we
found that it can identify about 99\%, 93\%, and 70\% (88\%) of stars,
AGN, and strong-candidate (plus weak-candidate) compact objects,
respectively. The corresponding probabilities of false identification,
defined as the ratio of the number of false identifications to the
total number of identifications, are about 1\%, 3\%, and 18\% (32\%),
respectively.

\subsection{Properties of Candidate Sources}
\label{sec:srcuid}

The results of applying the source type classification method outlined
in Section~\ref{sec:clsscrit} to the unidentified sources are given in
Tables~\ref{tbl:identifiedsrc} and \ref{tbl:dercat}. For very few
sources, our classification also used specific source properties (such
as dips) reported in the literature. The relevant references have been
included in Table~\ref{tbl:dercat}. We obtained 644 candidate
stars. For candidate AGN, we differentiated the 84 coincident with
(extended) galaxies (G) from the rest (1292). For compact objects, 202
are strong candidates (``CO'' in Tables~\ref{tbl:identifiedsrc} and
\ref{tbl:dercat}), 320 are candidate non-nuclear extragalactic sources
(XGSs), and 151 are candidate Galactic-plane sources (GPSs). These
sources do not include the 16 SNR, the 19 ``Mixed'' sources and the
100 candidate ULXs in Tables~\ref{tbl:identifiedsrc} and
\ref{tbl:dercat}. Among the 100 candidate ULXs, 11 were identified in
this work. They were selected under the conditions of the maximum
0.2--12.0 keV luminosity above 10$^{39}$ erg~s$^{-1}$ (6 above
2$\times$10$^{39}$ erg~s$^{-1}$), no optical counterparts within
2$\arcsec$ of the X-ray position from the USNO-B1.0 Catalog or the
SDSS, and $\alpha/R_{\rm 25}\le1.0$ (source \#106200 has
$\alpha/R_{\rm 25}=1.99$ and was included considering its large
variability ($V_{\rm var}$(0.2--4.5 keV)=10.9). In
Table~\ref{tbl:dercat}, we list the characteristics (being soft,
highly variable in X-rays, etc.) of the candidate compact objects and
ULXs that differentiate them from AGN.

\begin{figure*} 
\centering
\includegraphics{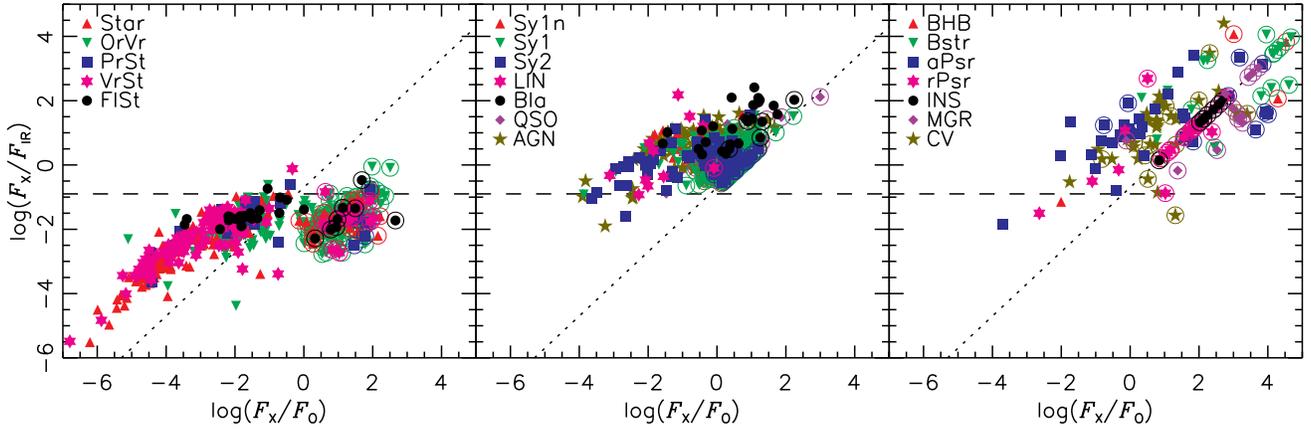}
\caption{The X-ray-to-IR versus X-ray-to-optical flux ratios of identified sources. The dotted and dashed lines are plotted for reference. Data points circled have no optical (to the right of the dotted line) and/or IR (to the left of the dotted line) counterparts found, and their optical/IR fluxes were calculated using detection limits (see text).  \label{fig:xrayoirind}}
\end{figure*}

\begin{figure*}
\subfigure[Stars]{
\includegraphics[width=0.32\textwidth]{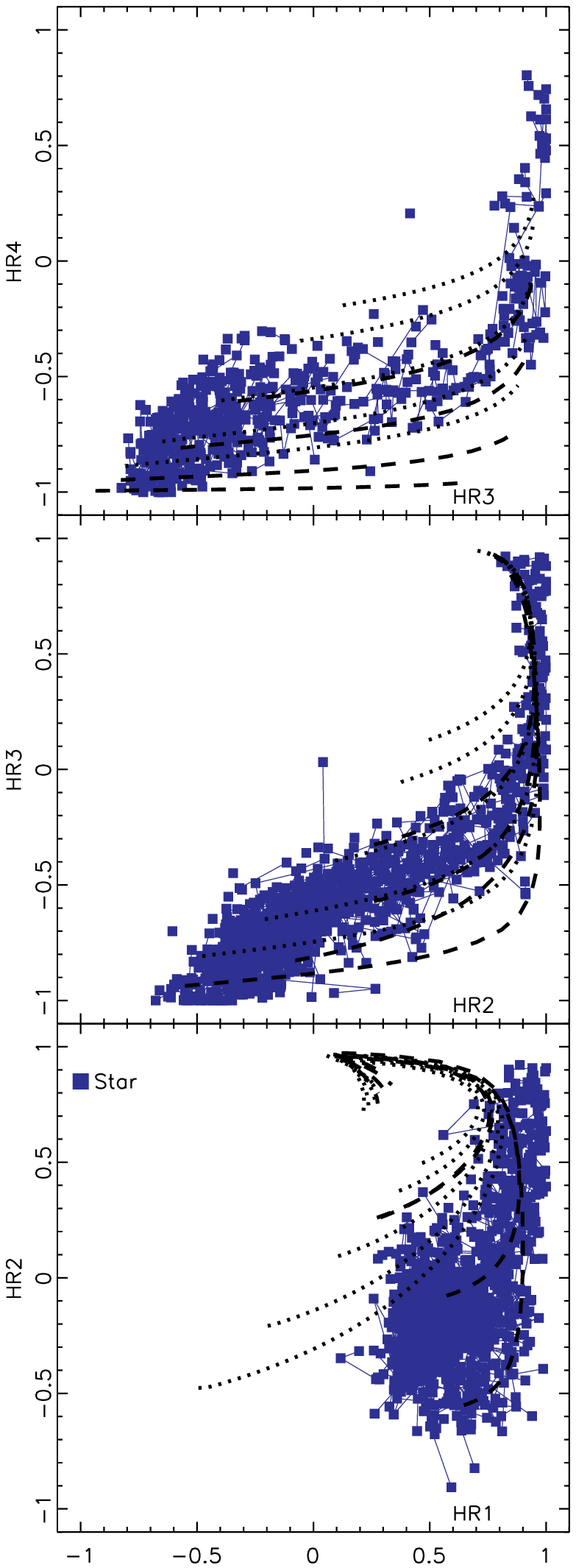}
\label{fig:starcolorcan}
}
\subfigure[AGN]{
\includegraphics[width=0.32\textwidth]{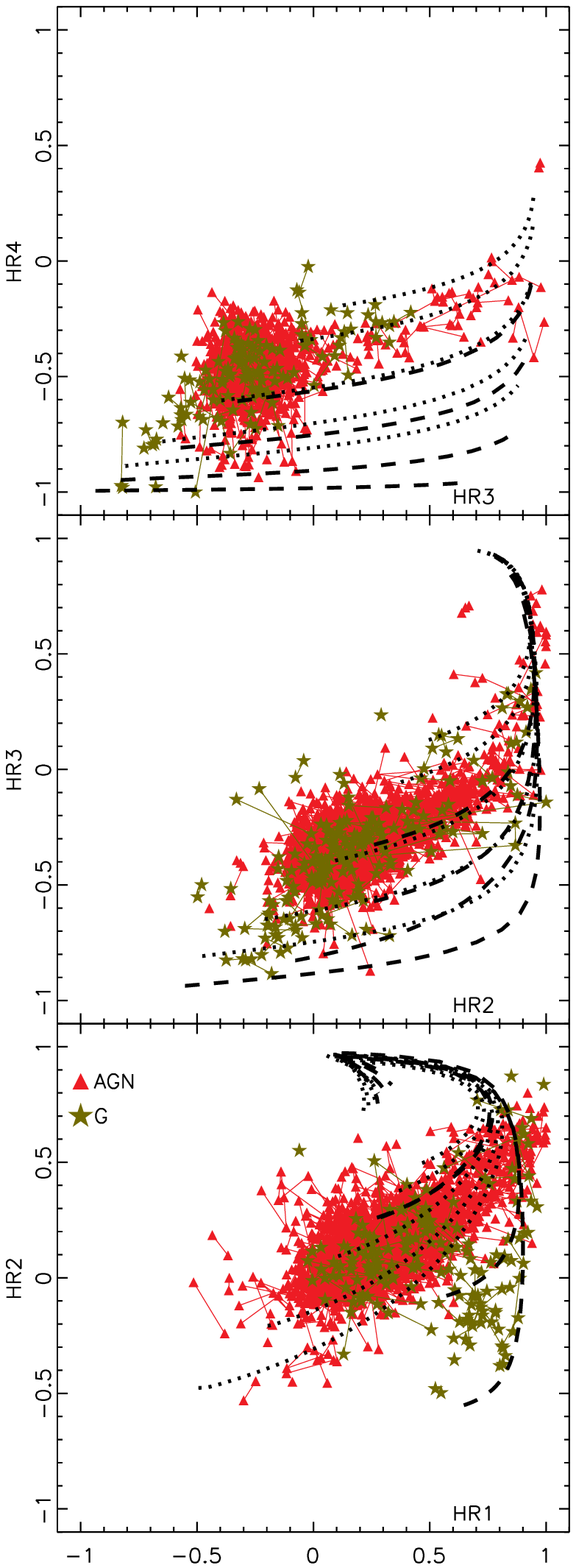}
\label{fig:agncolorcan}
}
\subfigure[Compact objects]{
\includegraphics[width=0.32\textwidth]{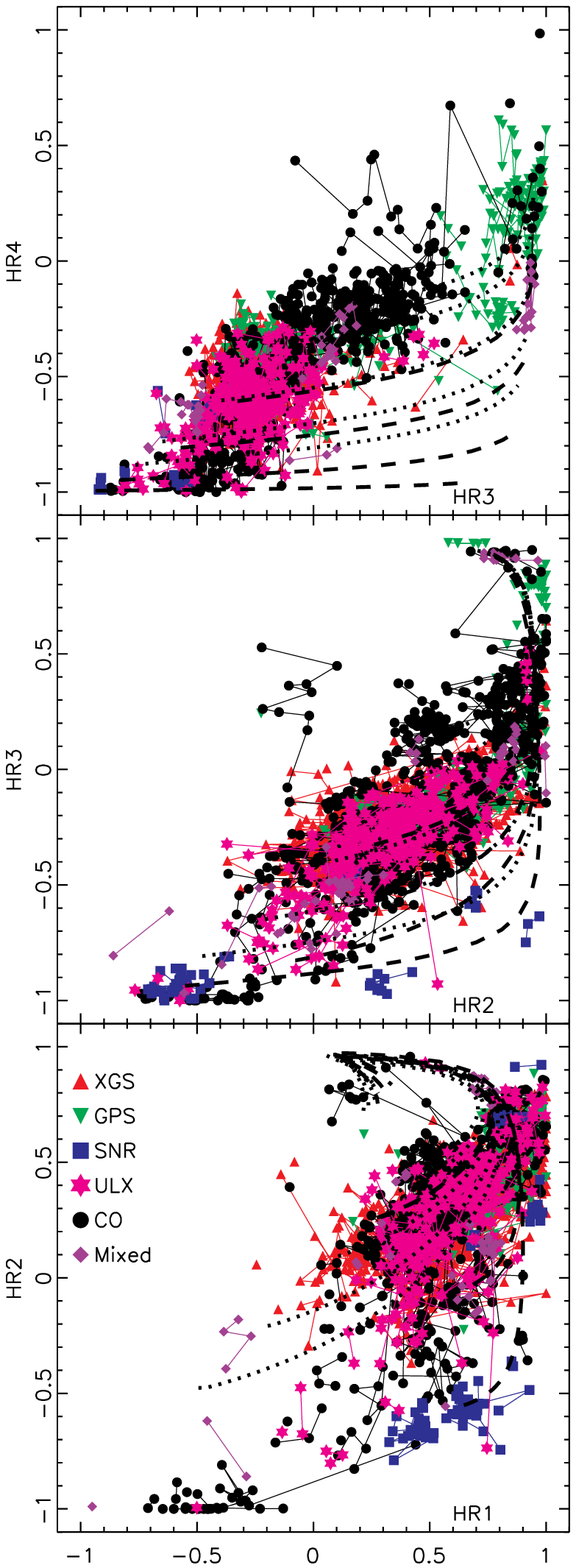}
\label{fig:cocolorcan}
}
\caption{The X-ray color-color diagrams, the same as Figure~\ref{fig:xraycolor} but for candidate sources. \label{fig:xraycolorcan}}
\end{figure*}

\begin{figure*}
\subfigure[Stars]{
\includegraphics[width=0.32\textwidth]{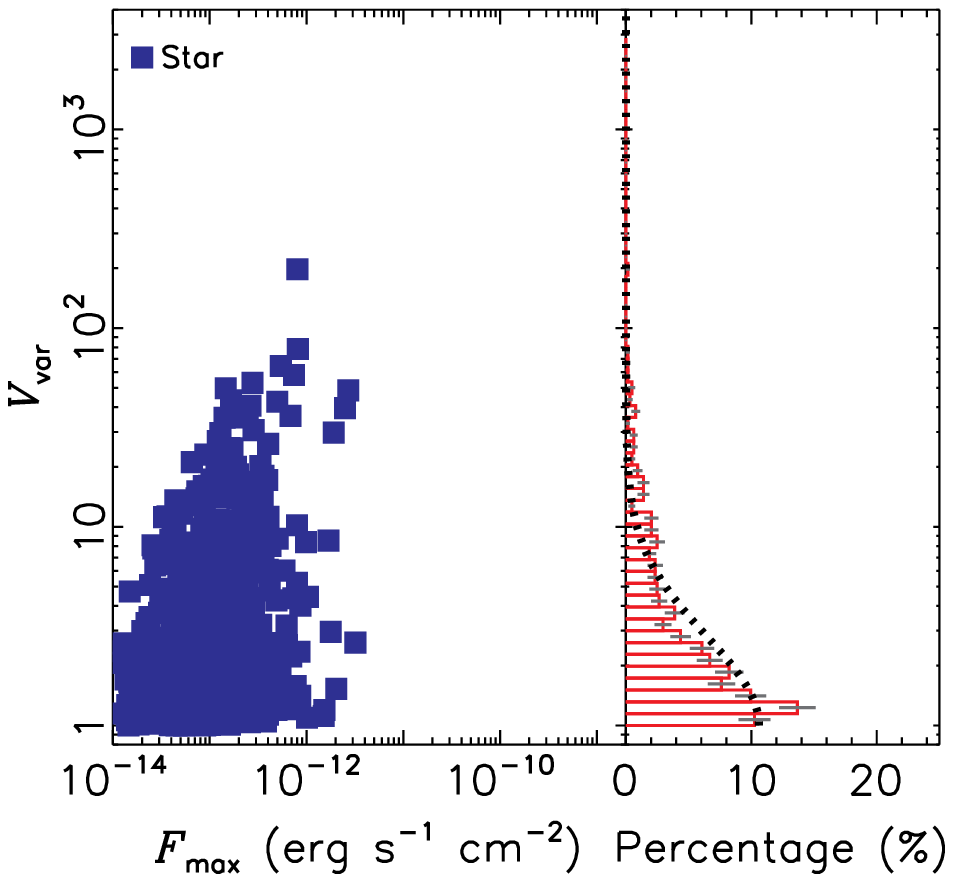}
\label{fig:starfvarcan}
}
\subfigure[AGN]{
\includegraphics[width=0.32\textwidth]{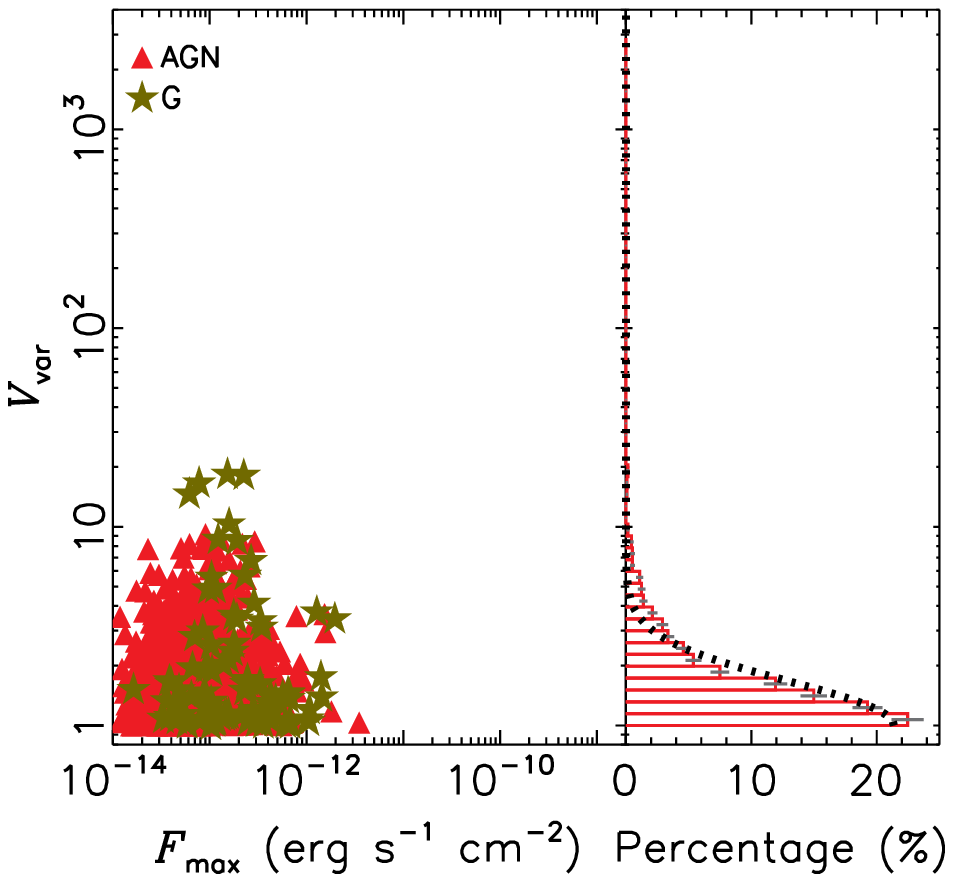}
\label{fig:agnfvarcan}
}
\subfigure[Compact objects]{
\includegraphics[width=0.32\textwidth]{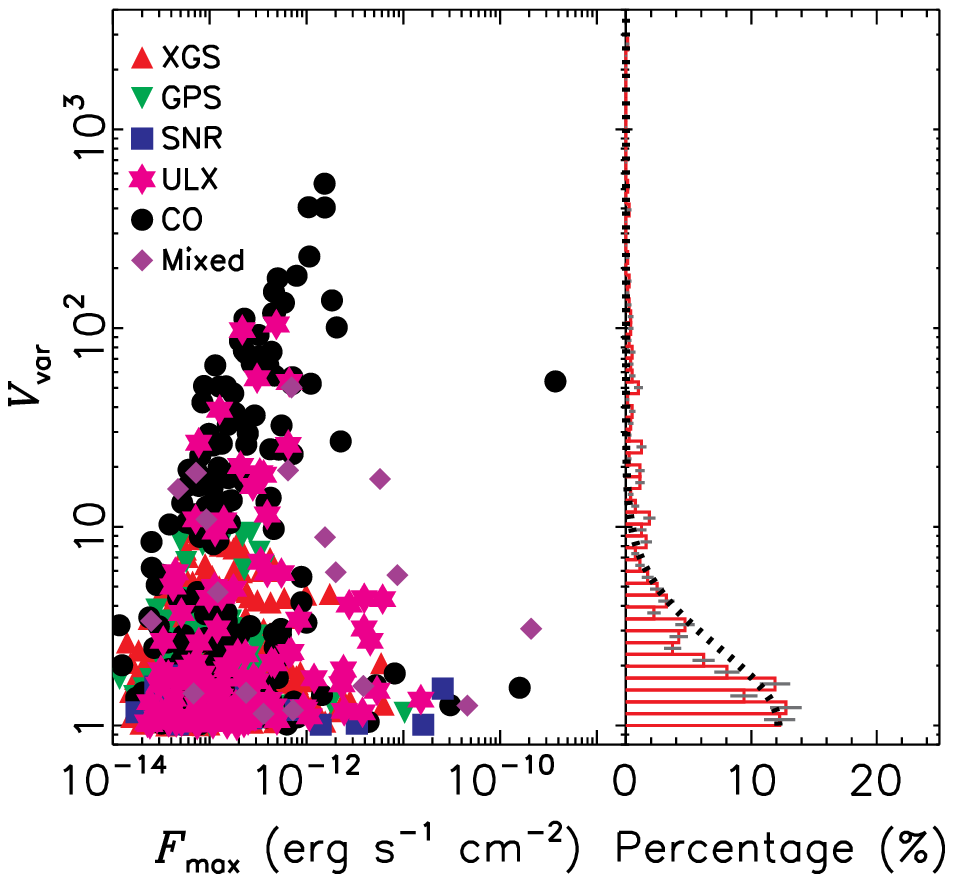}
\label{fig:cofvarcan}
}
\caption{The X-ray flux variation factor versus the maximum flux in 0.2--4.5 keV and the distribution of the flux variation factor, the same as Figure~\ref{fig:xrayfvar} but for candidate sources.\label{fig:xrayfvarcan}}
\end{figure*}

\begin{figure*} 
\centering
\includegraphics{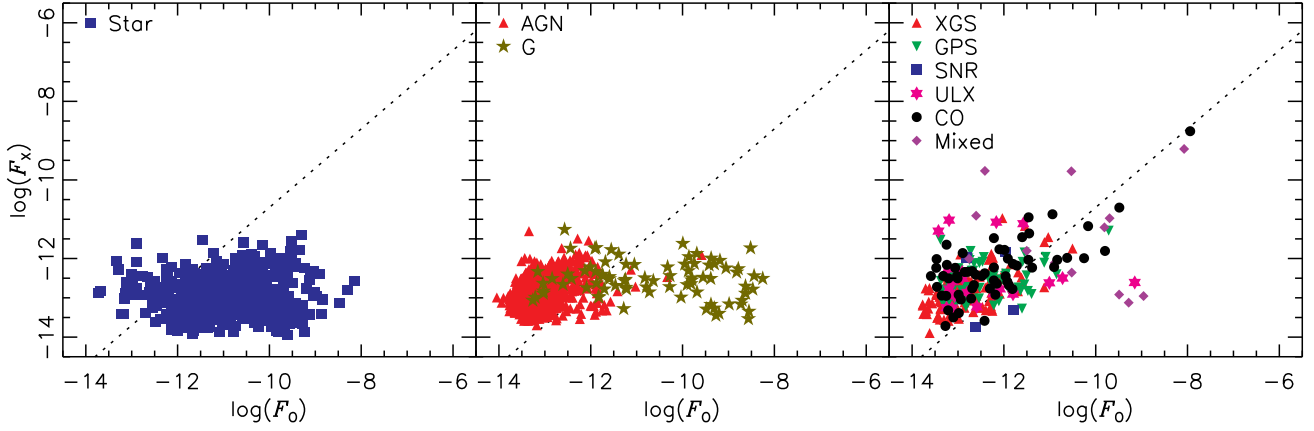}
\caption{The X-ray (0.2--12.0 keV; maximum) versus optical (USNO-B1.0 $R2$-band) fluxes, the same as Figure~\ref{fig:xrayofluxind} but for candidate sources.\label{fig:xrayofluxindcan}}
\end{figure*}

\begin{figure*} 
\centering
\includegraphics{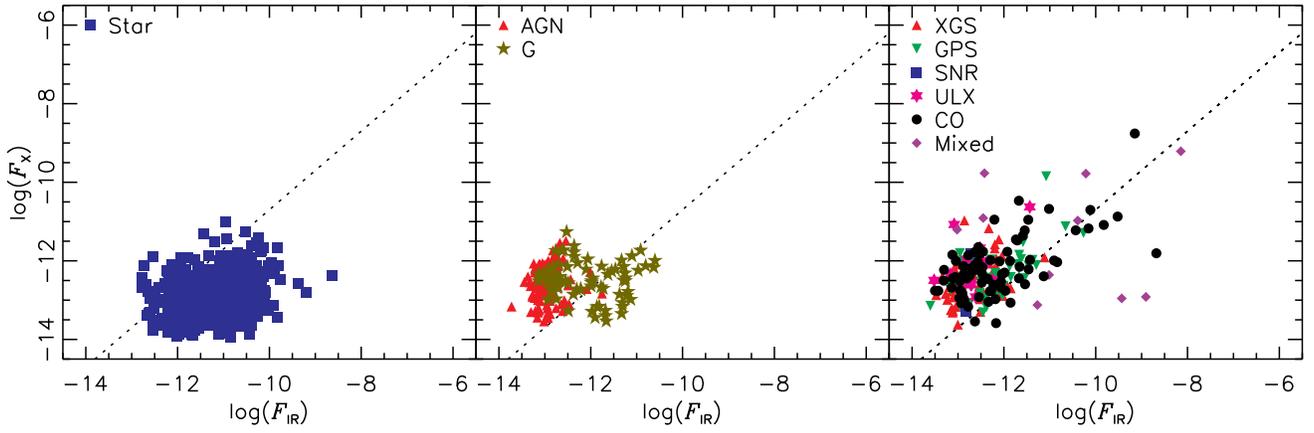}
\caption{The X-ray (0.2--12.0 keV; maximum) versus IR (2MASS $K_{\rm s}$-band) fluxes, the same as Figure~\ref{fig:xrayirfluxind} but for candidate sources.\label{fig:xrayirfluxindcan}}
\end{figure*}

\begin{figure*} 
\centering
\includegraphics{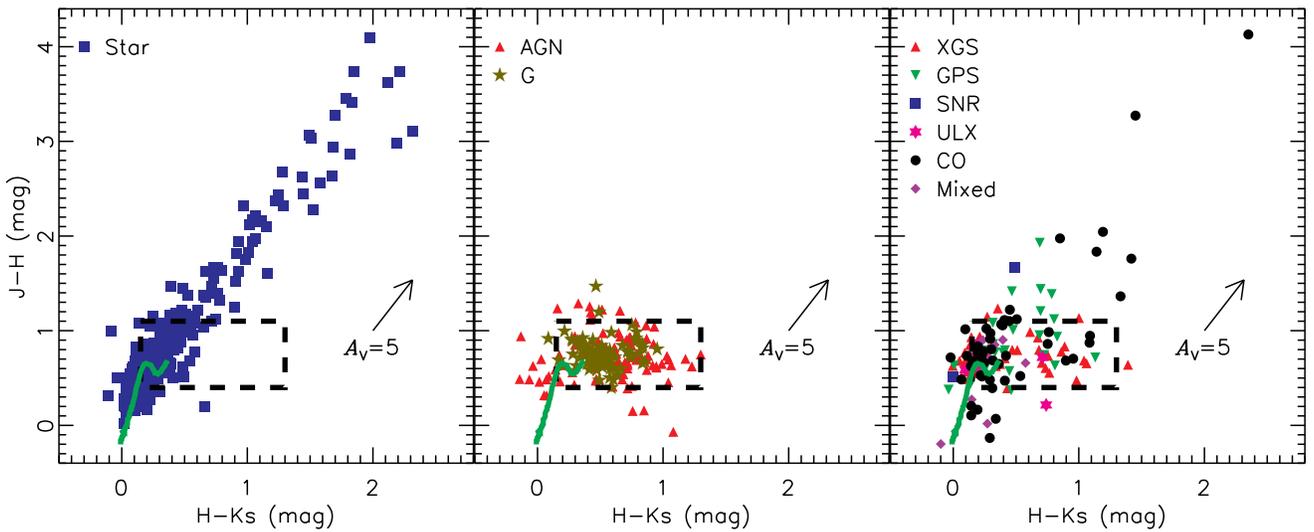}
\caption{The 2MASS color-color diagram, the same as Figure~\ref{fig:2masscolorind} but for candidate sources. Two stars, \#146483 ($J-H=4.92$, $H-K_{\rm s}=3.14$) and \#146372 ($J-H=5.74$, $H-K_{\rm s}=3.91$), are outside the plotting range. \label{fig:2masscolorindcan}}
\end{figure*}

\begin{figure*} 
\centering
\includegraphics{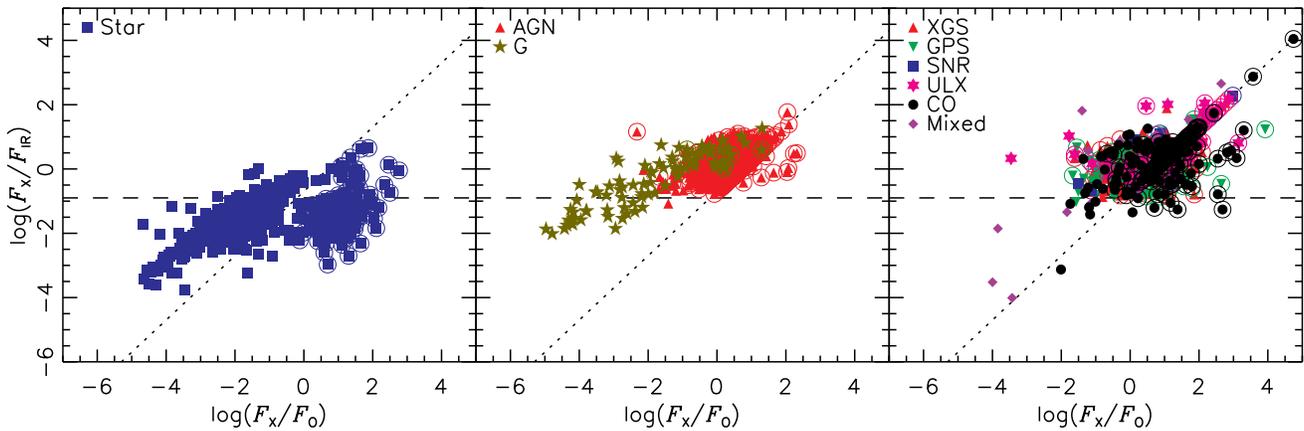}
\caption{The X-ray-to-IR versus X-ray-to-optical flux ratios, the same as Figure~\ref{fig:xrayoirind} but for candidate sources.\label{fig:xrayoirindcan}}
\end{figure*}

Figures~\ref{fig:xraycolorcan}--\ref{fig:xrayoirindcan} show the X-ray
color-color diagrams, long-term X-ray variability, the X-ray versus
optical fluxes, the X-ray versus IR fluxes, IR color-color diagrams,
and the X-ray-to-IR versus X-ray-to-optical flux ratios, respectively,
for these candidate sources. They correspond to
Figures~\ref{fig:xraycolor}, \ref{fig:xrayfvar} and
\ref{fig:xrayofluxind}--\ref{fig:xrayoirind} for identified sources,
respectively. The candidate and the identified sources generally show
similar properties in these plots. There are some subtle
differences. Comparing
Figures~\ref{fig:xrayofluxindcan}--\ref{fig:xrayirfluxindcan} with
Figures~\ref{fig:xrayofluxind}--\ref{fig:xrayirfluxind}, we see that
the candidate sources on the whole are fainter, mostly with the
maxiaml 0.2--12.0 keV flux $<$$10^{-12}$ erg~s$^{-1}$cm$^{-2}$. This can
be explained, as brighter X-ray sources tend to have brighter optical
counterparts, which makes the identification, typically through
optical measurements, easier.

Comparing Figures~\ref{fig:xrayoirind} and \ref{fig:xrayoirindcan}, we
see that there are more candidate stars than identified ones with
$\log(F_{\rm X}/F_{\rm IR})$$>$$-0.9$. This may be because some of
them are faint in IR and optical, hense hard to identify, while X-ray
flares are more effective to identify these sources. Sources \#173834
and \#6007 are two extreme cases. We measured $\log(F_{\rm X}/F_{\rm
  IR})=0.63$ for source \#173834, due to a large flare lasting
$\sim$20 ks in observation 0200450401. It was not detected in five
other observations, and we obtained a long-term flux variation factor
$V_{\rm var}$(0.2--4.5 keV)$>$53. Our classification as a star is also
supported by appearing within the Cygnus OB2 star forming
region. Source \#6007 has $\log(F_{\rm X}/F_{\rm IR})=0.65$, due to a
large flare lasting $\sim$8 ks in observation 0505720501. It appears
within M 31, and it was also classifed as a foreground star by
\citet{stpiha2008} due to a (much weaker) flare in an earlier
observation. We measured $V_{\rm var}$(0.2--4.5 keV)$>$198. Sources
with extreme properties warrant future investigation.

The 84 candidate AGN that are coincident with the centers of the
extended galaxies on the whole have brighter optical and IR
counterparts than other candidate AGN
(Figures~\ref{fig:xrayofluxindcan}--\ref{fig:xrayirfluxindcan}). They
also have lower X-ray-to-IR and X-ray-to-optical flux ratios, which
separate them from other candidate AGN in
Figure~\ref{fig:xrayoirindcan}, probably indicating large
contamination of IR and optical counterpart fluxes from the
starlight. In the X-ray color-color diagrams in
Figure~\ref{fig:xraycolorcan}, some of them are also separated from
other candidate AGN and are more in the region occupied by stars,
indicating that their X-ray emission may be dominated by the hot gas
inside the host galaxies, as often seen in nonactive early-type
galaxies \citep{fa1989}.

The 16 SNRs are remarkably similiar to each other. They all have X-ray
colors similar to stars (Figure~\ref{fig:xraycolorcan}). Their X-ray
emission is probably due to expanding SNR shock heating the
surrounding interstellar medium \citep[e.g.,][]{itma1989} and can be
fitted with a bremsstrahlung model with temperatures $<$1 keV
\citep{padula2000}. They either (for most cases) have no IR or optical
counterparts found from the USNO-B1.0 Catalog and the 2MASS PSC, or
have counterparts at large separations from the X-ray positions
($>$2$\arcsec$), thus probably spurious. They are very steady, with
the 0.2--4.5 keV flux variation factors all $<$1.8 (the median is
1.19, Figure~\ref{fig:xrayfvarind}) and the variability significance
$S_{\rm var}$ all $<$4.

\begin{figure} 
\centering
\includegraphics[width=0.35\textwidth]{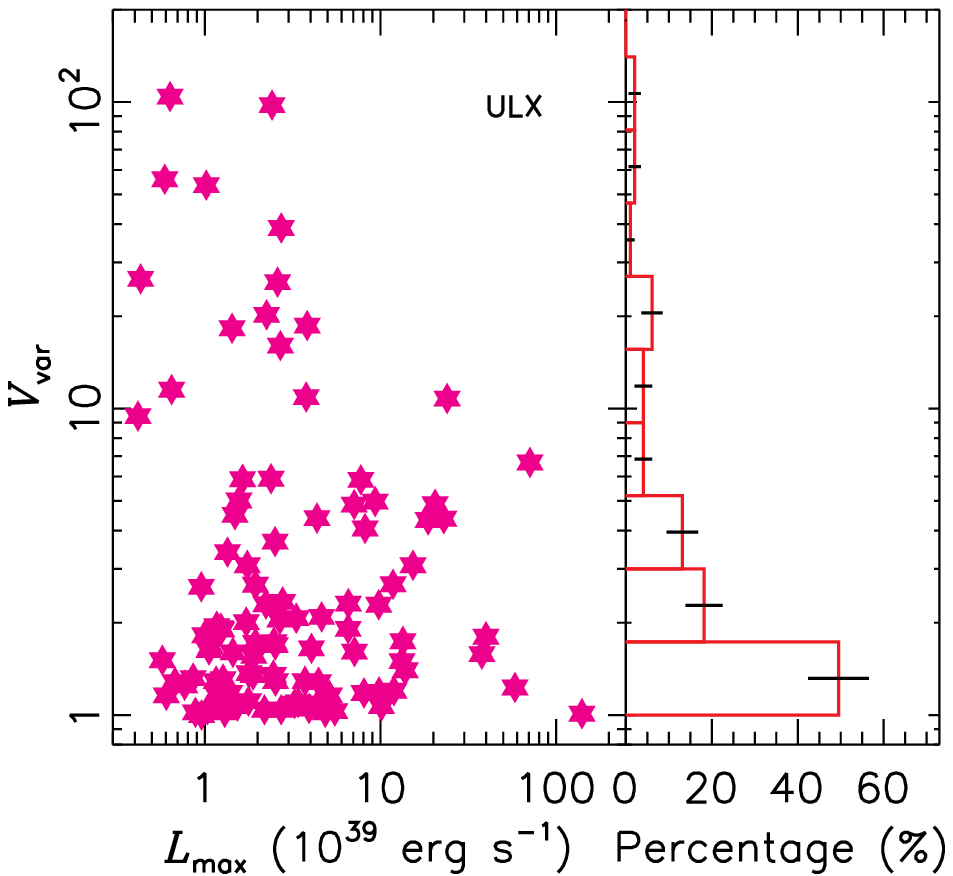}
\caption{The X-ray flux variation factor (0.2--4.5 keV) versus the maximum luminosity (0.2--12.0 keV) and the distribution of the flux variation factor for candidate ULXs.\label{fig:ulxxvar}}
\end{figure}

For the 100 candidate ULXs, most of them are in the region occupied by
AGN in the X-ray color-color diagrams in
Figure~\ref{fig:xraycolorcan}, while nearly 30\% of them have
detections entering the soft regions (HR1$<$$-0.4$, HR2$<$$-0.5$,
HR3$<$$-0.7$, or HR4$<$$-0.8$) where AGN are hardly seen. The ULX
sample are often contaminated by background AGN \citep{logu2006}, and
we expect that some of these 100 candidate ULXs are in fact
AGN. Figure~\ref{fig:ulxxvar} plots the X-ray flux variation factor
(0.2--4.5 keV) versus the maximum luminosity (0.2--12.0 keV) and the
distribution of the flux variation factor for these candidate
ULXs. There are 13 candidates (found from the literature) with the
maximum 0.2--12.0 keV luminosity slightly less than 10$^{39}$
erg~s$^{-1}$, the lowest limit generally used to define ULXs. This
discrepancy can be explained by various factors, such as source
variability and different source distances used in the literature from
us (we mostly refer the distances to \citet{libr2005} and
\citet{li2011}). About 15\% of these 100 candidate ULXs have $V_{\rm
  var}$(0.2--4.5 keV)$>$10, but most of them have the maximum
0.2--12.0 keV luminosity lower than 4$\times$10$^{39}$ erg~s$^{-1}$.
We obtained a median of 1.73 for the 0.2--4.5 keV flux variation
factors, which is slightly larger than that of AGN but smaller than
those of various types of accreting compact objects
(Section~\ref{sec:ltxvarid}). Although we do not have enough BHBs to
obtain a meaningful value of their average variation factor, we expect
it to be much larger than that of these candidate ULXs, considering
that most BHBs are transients \citep{mcre2006}.

For the 202 strong candidate compact objects, they are spread across
the X-ray color-color diagrams (Figure~\ref{fig:xraycolorcan}); they
include some supersoft X-ray sources and some hard sources which are
probably accretion-powered X-ray pulsars. The weak candidate compact
objects in the Galactic plane are mostly highly absorbed, occupying
the upper right corner in the HR1-HR2 diagram.

Several special objects have very low X-ray-to-IR flux ratios in the
right panel in Figure~\ref{fig:xrayoirindcan} though they probably
contain compact objects. The four candidate symbiotic stars, i.e.,
\object{AG Draconis} (\#143273), \object{4U 1700+24} (\#152155),
\object{Z And} (\#189311), and \object{SMC Symbiotic Star 3}
(\#194075), all have $\log(F_{\rm X}/F_{\rm IR})$$<$$-1.0$ (they are
included in the ``Mixed'' class). One supergiant fast X-ray transient
(SFXT), i.e., \object{IGR J17544-2619} (\#158580), has $\log(F_{\rm
  X}/F_{\rm IR})$$=$$-1.0$ \citep[it is included in the ``CO''
  class,][]{goooku2004}. There is another SFXT \object{IGR
  J11215-5952} (\#206717), which has been included in the class of
accretion-powered X-ray pulsars due to the detection of the spin
period \citep{sirome2007} and has $\log(F_{\rm X}/F_{\rm IR})$ only
$-0.1$. We also detected flares lasting $\sim$1 h from these
two SFXTs. Their flares, however, show complicated structure, in
contrast with the typical fast rise and slow decay of stellar X-ray
flares. Observations of very hard spectra, short pulsation periods and
extreme long-term variability as in \object{IGR J11215-5952}
\citep{sirome2007} can help to differentiate them from stars.

\section{DISCUSSION AND CONCLUSIONS}
\label{sec:conclusion}
We have systematically studied the properties of 4330 point sources
from the 2XMMi-DR3 catalog. These sources represent a high-quality
X-ray source sample, each with multiple observations and at least one
detection with S/N$\ge$20. They have very high astrometric accuracy
(99\% with the positional error $<$1$\arcsec$). For about one third of
them we have obtained reliable source types from the literature. They
correspond to various types of stars, AGN and compact object systems
containing WDs, NSs, and stellar-mass BHs. We have studied their
properties in terms of the X-ray spectral shape, X-ray variability,
and multi-wavelength cross-correlation. We find that 99\% of stars can
be separated from other sources using the X-ray-to-IR flux ratio and
the X-ray flare. Stars also occupy well defined regions in the X-ray
and IR color-color diagrams where sources of other types, especially
AGN, seldom visit. Comparing AGN and compact objects, the X-ray
spectra of AGN are remarkably similar to each other, with $\Gamma_{\rm
  PL}$ concentrated around 1.91$\pm$0.31, and their long-term flux
variation factors have a median of 1.48 and 98.5\% less than 10, while
70\% of compact object systems can be very soft or hard, highly
variable in X-rays, and/or have very large X-ray-to-IR flux ratios,
separating them from AGN. Using the above results, we have derived a
source type classification method to classify the rest of the sources
in our sample.

In our source type classification method, we select sources with
$\log(F_{\rm X}/F_{\rm IR})$$<$$-0.9$ and/or X-ray flares to be stars
unless they have HR1$\le$0.3 and/or are coincident with the centers of
extended optical/IR sources, presumably galaxies. Due to the
requirement that the source has at least one detection with S/N$\ge$20
in our source selection criteria, our sources have maximum 0.2--12.0 keV
flux $10^{-14}$ to $10^{-9}$ erg~s$^{-1}$~cm$^{-2}$, corresponding to
relativelly bright sources in the 2XMMi-DR3 catalog. For too faint
sources when their IR counterparts fall well below the 2MASS detection
limit, it becomes hard to estimate their $\log(F_{\rm X}/F_{\rm IR})$.
For them, the X-ray color and the flare (about 20\% of stars have
flares detected by us) will be more effective to separate stars from
other types of sources than the X-ray-to-IR flux ratio. To separate
compact object systems from AGN, we rely on X-ray colors, X-ray
long-term variability and the X-ray-to-IR flux ratio. There are many
other strong indicators of compact object systems, such as dips,
eclipses, pulsations (coherent or noncoherent) and bursts, but they
normally require high-quality data and detailed analysis.

We find 644 candidate stars, 1376 candidate AGN and 202 candidate
compact object systems, whose spurious probabilities are estimated to
be about 1\%, 3\% and 18\%, respectively, based on the test of our
classification method on the identified sources. There are still 320
appearing to be associated with nearby galaxies and 151 appearing in
the Galactic plane, which are probably compact object systems or
background AGN. Our source sample also contain 100 candidate ULXs,
which include 11 new ones. With long-term flux variation factors
having a median of only 1.73 and 85\% less than 10, they are much less
variable than other accreting compact objects. In this work, we have
also found a few tens of sources that show interesting properties but
are poorly studied in the literature. We will present our detailed
study of these sources in a companion paper.

There are other studies using multi-wavelength cross-correlation to
classify X-ray sources \citep[See, e.g., ][and the references
  therein]{haru2009}. \citet{haru2009} studied the cross-correlation
between the RASS Bright Source Catalog (RASS BSC) and the 2MASS PSC
and showed that stars are well separated from other types of sources
in terms of the X-ray-to-IR flux ratio and the IR color $J-K_{\rm
  s}$. As sources in the RASS BSC mostly have the 0.1--2.4 keV flux
$>$$10^{-12}$ erg~s$^{-1}$~cm$^{-2}$ \citep{voasbo1999}, our sample
extends the fluxes by two orders of magnitude below theirs. Although
we also find that stars generally have lower X-ray-to-IR flux ratios
than other types of sources, their result that stars have $J-K_{\rm
  s}\le1.1$ is not well supported by our source sample. Our identified
and candidate stars have $J-K_{\rm s}$ spanning from about 0.0 to 8.0,
and about 30\% and 20\% of them have $J-K_{\rm s}\ge1.1$,
respectively. The high value of the $J-K_{\rm s}$ color for these
stars is probably caused by large extinction. The lack of these highly
absorbed sources in the RASS BSC is easy to explain, as the energy
pass of {\it ROSAT} is in the very soft X-rays (0.1--2.4 keV). Because
of these highly absorbed sources, we choose to use the $K_{\rm
  s}$-band flux to calculate the X-ray-to-IR flux ratio instead of the
$J_{\rm s}$-band flux used in \citet{haru2009}. This is also why stars
in our sample are not well separated from other types of sources in
terms of the X-ray-to-optical flux ratio, though this is observed to
be the case in many studies \citep[e.g.,][]{voasbo1999}.

\appendix
\section{Estimate of the systematic errors of X-ray fluxes and hardness ratios}
\label{sec:fhrerr}
The 2XMMi-DR3 catalog only gives the statistical errors for the X-ray
fluxes. Their systematic errors are estimated as follows. We first
considered the systematic difference of the fluxes between pn and
MOS. The MOS flux is the average of the fluxes of MOS1 and MOS2 (they
have very similar responses) weighted by the errors. Using detections
with both the pn and MOS fluxes and at least one of them above 5
$\sigma$ from our source sample, we determined the systematic errors
that make 31.7\% (corresponding to 1-$\sigma$ difference) of
detections with the pn and MOS flux difference larger than the total
error, i.e.,
\begin{equation}
|F_{\rm pn}-F_{\rm MOS}|\ge \nonumber [(r_{\rm pn}\times F_{\rm pn})^2+(r_{\rm MOS}\times F_{\rm MOS})^2 +(\sigma_{\rm pn})^2+(\sigma_{\rm MOS})^2]^{1/2},
\end{equation}
where $F_{\rm pn}$ and $F_{\rm MOS}$ are the pn and MOS fluxes, with
the 1-$\sigma$ statistical errors $\sigma_{\rm pn}$ and $\sigma_{\rm
  MOS}$ and fractional systematic errors $r_{\rm pn}$ and $r_{\rm
  pn}$, respectively. Assuming $r_{\rm pn}=r_{\rm MOS}$, we estimated
$r_{\rm pn}=r_{\rm MOS}=0.052$ and 0.076 for energy bands 14 and 8,
respectively. There are also systematic errors due to different
pointing offsets, data modes, etc. for different observations and the
approximation of the PL spectrum, and they are hard to estimate. In
the end, we just assumed the fractional systematic errors of the EPIC
fluxes in energy bands 14 and 8 to be $r=\sqrt{2}r_{\rm pn}=0.074$ and
0.107, respectively.

The systematic differences of the hardness ratios between the pn and
MOS can be estimated in a similar way. Using detections with both the
pn and MOS hardness ratios and at least one of their statistical
errors less than 0.1, we determined the systematic errors that make
31.7\% of detections with the difference of the pn and MOS hardness
ratios larger than the total error, i.e.,
\begin{equation}
|{\rm HR}i_{\rm pn}-{\rm HR}i_{\rm MOS}|\ge [(s_{i,\rm pn})^2+(s_{i,{\rm MOS}})^2+(\sigma_{i, \rm pn})^2+(\sigma_{i, \rm MOS})^2]^{1/2},
\end{equation}
where ${\rm HR}i_{\rm pn}$ and ${\rm HR}i_{\rm MOS}$ are the hardness
ratios defined as in Equation~\ref{eq:hr} using the pn and MOS fluxes,
respectively, with the statistical errors $\sigma_{i, \rm pn}$ and
$\sigma_{i, \rm MOS}$ and the absolute systematic errors $s_{i,\rm
  pn}$ and $s_{i,\rm MOS}$, respectively. Assuming $s_{i,\rm
  pn}=s_{i,{\rm MOS}}$, we obtained $s_{i,\rm pn}=s_{i,{\rm
    MOS}}=0.036$, 0.041, 0.014, and 0.030 for $i$=1 to 4,
respectively. In comparison, the hardness ratios in our definition
assume values between $-1$ and 1.

\section{Search for stellar X-ray flares}
\label{sec:flaresearch}
We searched for stellar X-ray flares as follows. To minimize the risk
of false detection of flares due to flaring background, we excluded
data in the period of flaring background in the search. We first
located the data point $i$ at time $t_i$ which has the maximum
weighted count rate $R_i$ calculated from its nearby three data points
$i$, $i+1$, and $i+2$. The weighted error of $R_i$ is denoted as
$R_{i,\rm err}$. We then calculated the weighted background count rate
$b_{i}$ and weighted standard deviation $b_{i,\rm err}$ using data in
the background time windows [$t_i-40$ks, $t_i-t_{\rm win}$] and
[$t_i+2t_{\rm win}$, $t_i+40$ks], where $t_{\rm win}$ is equal to 10
ks or one fifth of the observation duration for observations lasting
$<$50 ks. We identified a flare candidate at $t_i$ if $R_i-2R_{i,\rm
  err}>b_{i}+2b_{i,\rm err}$ and $R_i/b_i>5$. This process continued,
but using only data 40 ks outside the previously identified flare
candidates. The test on identified stars (Section \ref{sec:srcclslit})
showed that the above procedure can successfully identify most
flares. Some flares (about 10\%) are missed mostly because they are
very weak or the observations are short. A few non-stellar flares from
sources with large fast variability (such as periodic oscillations in
AM Herculis objects) were also picked up. Their profiles are generally
very complicated and very different from the typical profiles of smooth
fast rise and slow decay of stellar flares, and we discarded them
based on visual inspection. In Table~\ref{tbl:dercat} we give the
maximum ratio $R_i/b_i$ and the corresponding observation for sources
with stellar flares found. For sources that have no flares found using
the above search procedure but have some flares from our visual
inspection, this ratio is set to be one.

\clearpage

\tabletypesize{\tiny}
\setlength{\tabcolsep}{0.005in}
\begin{deluxetable*}{cccccccccccccccccccccccc}
\tablecaption{Source Detections and Simple Fits with an Absorbed PL \label{tbl:A1}}
\tablewidth{0pt}
\tablehead{SRCID & OBSID & SRCIDDet & R1 &R1\_err & R2 &R2\_err & R3 &R3\_err & R4 &R4\_err & R5 &R5\_err & $\chi^2_\nu$ & dof
  & $N_{\rm H}$ & $N_{\rm H, le}$ & $N_{\rm H, ue}$
  & $\Gamma_{\rm PL}$ & $\Gamma_{\rm PL, le}$ & $\Gamma_{\rm PL, ue}$ 
  & $F_{\rm PL}$ & $F_{\rm PL, le}$ & $F_{\rm PL, ue}$\\
(1) &(2) &(3) &(4) &(5) &(6) &(7) &(8) &(9) &(10) &(11) &(12) &(13) &(14) &(15) &(16) &(17) &(18) &(19) &(20) &(21) &(22) & (23) & (24)
}
\startdata
837 &0101040101 &837 &1.67e-02 &7.46e-04 &1.40e-02 &6.82e-04 &7.56e-03 &5.19e-04 &2.49e-03 &3.11e-04 &4.99e-04 &2.89e-04 &1.4 &6 &0.00 &0.00 &0.03 &2.92 &2.79 &3.23 &3.90e-13 &3.03e-13 &4.37e-13 \\
837 &0306870101 &837 &1.03e-02 &3.10e-04 &9.23e-03 &3.08e-04 &5.68e-03 &2.58e-04 &1.92e-03 &1.60e-04 &5.22e-04 &1.63e-04 &2.0 &6 &0.00 &0.00 &0.02 &2.72 &2.61 &2.95 &2.54e-13 &2.07e-13 &2.85e-13 \\
837 &0510010701 &837 &8.55e-03 &4.93e-04 &7.10e-03 &4.96e-04 &4.42e-03 &4.83e-04 &1.58e-03 &3.18e-04 &1.01e-04 &2.31e-04 &0.9 &6 &0.00 &0.00 &0.03 &2.69 &2.50 &3.13 &1.69e-13 &1.35e-13 &1.94e-13 \\
938 &0101040101 &938 &1.78e-03 &1.37e-04 &2.23e-03 &1.57e-04 &2.55e-03 &1.90e-04 &9.24e-04 &1.27e-04 &5.51e-04 &1.33e-04 &1.6 &10 &0.00 &0.00 &0.03 &2.04 &1.90 &2.33 &6.75e-14 &5.74e-14 &7.93e-14 \\
938 &0306870101 &938 &1.12e-03 &1.05e-04 &1.92e-03 &1.39e-04 &2.26e-03 &1.57e-04 &1.44e-03 &1.30e-04 &4.61e-04 &1.32e-04 &0.9 &6 &0.00 &0.00 &0.04 &1.72 &1.58 &1.98 &8.64e-14 &7.02e-14 &1.01e-13 \\
938 &0510010701 &938 &1.03e-03 &1.72e-04 &1.14e-03 &1.99e-04 &9.62e-04 &2.44e-04 &3.32e-04 &1.72e-04 &1.51e-04 &1.81e-04 &0.6 &6 &0.01 &0.00 &0.12 &2.42 &1.96 &3.58 &2.70e-14 &1.70e-14 &3.81e-14
\enddata 
\tablecomments{This table is published in its entirety in the electronic edition of the Astrohysical Journal. A portion is shown here for guidance regarding its form and content. Columns: (1) the unique source number that we used to refer the source, (2): observation ID, (3): the unique source number in the 2XMMi-DR3 catalog: (4)--(13): the MOS1-medium-filter equivalent counts rates in five basic energy bands and the corresponding errors (in units of cts/s), (14): the reduced $\chi^2$ of the fit with an absorbed PL, (15) degrees of freedom, (16)--(18): the best-fitting $N_{\rm H}$ and lower and upper bounds (at a 90\% confidence level, in units of 10$^{22}$ cm$^{-2}$), (19)--(21): the best-fitting PL photon index and lower and upper bounds, (22)--(24): the 0.2--12.0 keV flux from the PL fit and the lower and upper bounds (in units of erg cm$^{-2}$ s$^{-1}$). For a unique source, the detections are specified by OBSID and SRCIDDet, which are ``OBS\_ID'' and ``SRCID'' in the 2XMMi-DR3 catalog, respectively. SRCID is different from SRCIDDet only when several ``unique'' sources in the 2XMMi-DR3 catalog are rematched together (see Table 1). When the source has no detection in the 2XMMi-DR3 catalog for an observation, the corresonding SRCIDDet is empty, and the fit was not carried out. There are 17 detections not used in our study (thus not in this table too), because of large discrepancy between pn and MOS measurements or large extended radius used in the 2XMMi-DR3 catalog (causing fake large fluxes).}
\end{deluxetable*}

\setlength{\tabcolsep}{0.01in}
\begin{deluxetable*}{rlccccccccccccccccccc}
\centering
\tablecaption{General Source Properties and Source Classification \label{tbl:dercat}}
\tablewidth{0pt}
\tablehead{SRCID & \colhead{2XMMi-DR3} &\colhead{RAdeg}&\colhead{DEdeg} &\colhead{NObs} & \colhead{DObs} &\colhead{NDet} &\colhead{DDet} & \colhead{Type} &\colhead{TpCat}&\colhead{Ref} &\colhead{RefType} & \colhead{SrcChar}\tablenotemark{a} & \colhead{Rxo} & \colhead{Rxir} & \colhead{Vvar14} &  \colhead{Dxo} & \colhead{Dxir} & \colhead{RC3Name} & \colhead{RC3SepR}\\
(1) &\colhead{(2)} &(3) &(4) &(5) &(6) &(7) &(8) &(9) &(10) &(11) &(12) &(13) &(14) &(15)&(16) &(17) &(18) &(19)&(20)
}
\startdata
8085 &  2XMM J005412.9-373309 &  13.55411 & $-37.55272$ &  4 &  1794.51 &  4 &  1794.51 &    CO & B &\nodata&\nodata&   SVE & $ 1.26$ & $ 0.56$ &   36.33 &\nodata&\nodata&          NGC300 & 1.11  \\
100081 &  2XMM J120429.7+201858 & 181.12376 & $ 20.31638$ &  3 &   178.09 &  3 &   178.09 &   Sy2 & A &                 VV10 &               A/S2 &\nodata& $-2.06$ & $ 0.11$ &    2.04 &  0.6 &  0.8 &\nodata&      \\
101335 &  2XMM J121028.9+391748 & 182.62064 & $ 39.29689$ &  8 &  2169.02 &  8 &  2169.02 &    CO & B &\nodata&\nodata&   S & $ 0.70$ & $0.00$ &    2.63 &\nodata&\nodata&\nodata&      \\
107102 &  2XMM J123103.2+110648 & 187.76337 & $ 11.11348$ &  3 &   887.86 &  3 &   887.86 &    CO & B &\nodata&\nodata&    S & $ 0.31$ & $ 0.12$ &    2.11 &  3.6 &\nodata&\nodata&      \\
112394 &  2XMM J125048.6+410743 & 192.70256 & $ 41.12836$ &  3 &  1649.30 &  2 &  1615.12 &   ULX & B &\nodata&\nodata&    VE & $ 1.67$ & $ 0.97$ &   25.85 &\nodata&\nodata&         NGC4736 & 0.17 \\
250424 & 2XMMi J170058.4-461107 & 255.24352 & $-46.18554$ &  3 &   209.32 &  3 &   209.32 &  Bstr & A &  2009ApJ...699...60L &              Bstr &\nodata& $ 1.92$ & $ 1.22$ &    3.15 &\nodata&\nodata&\nodata&  
\enddata 
\tablecomments{This table is published in its entirety in the electronic edition of the Astrophysical Journal. A portion is shown here for guidance regarding its form and content. Columns are as follows. (1): 2XMM-DR3 Unique source index, (2): 2XMMi-DR3 Source designation, (3): Mean source ICRS right ascension (J2000), (4): Mean source ICRS declination (J2000), (5)--(6): Number and time span (in units of days, using the observation start time) of \xmm\ observations, (7)--(8): Number and time span (in units of days, using the observation start time) of detections from the 2XMMi-DR3 catalog, (9): Source type, (10) The source type category (A=identified, B=candidate), (11)--(12): Reference and the spectral type in the reference, (13): The source characteristics based on which we classify the source as a compact object system (instead of star or AGN), (14): X-ray-to-optical flux ratio logarithm $\log(F_{\rm X}/F_{\rm O})$, (15): X-ray-to-IR flux ratio logarithm $\log(F_{\rm X}/F_{\rm IR})$, (16): Flux variation factor in the 0.2-4.5 keV band, (17): USNO-B1.0 counterpart separation from the X-ray position (in units of arcsec), (18): 2MASS counterpart separation from the X-ray position (in units of arcsec),  (19): The source name of the RC3 match, (20): The ratio of the angular separation to the $D_{25}$ isophote elliptical radius.}
\tablenotetext{a}{ ``S'': soft (HR1$<$$-0.4$, HR2$<$$-0.5$, HR3$<$$-0.7$, or HR4$<$$-0.8$); ``H'': hard ($-0.1$$<$HR3$<$0.5 and $-0.25$$<$HR4$<$0.1); ``V'': highly variable $V_{\rm var}$(0.2--4.5 keV)$>$10; ``R'': high X-ray-to-IR flux ratio ($\log(F_{\rm X}/F_{\rm IR}$)$>$2.5); ``E'': non-nuclear extra-galactic;  ``G'': in the Galactic plane ($|b|$$<$10); ``L'': literature, which includes more source properties such as dips.}
\end{deluxetable*}

\end{document}